\documentclass[11pt,twocolumn,numbers]{article}

\usepackage{graphicx}
\usepackage{amssymb}
\usepackage{amsthm,amsmath,dsfont}
\usepackage{mathtools}

\usepackage{lineno}

\usepackage[colorlinks=true, citecolor=blue]{hyperref}
\usepackage{algorithm2e}
\usepackage{algcompatible}
\usepackage{wrapfig}
\usepackage[numbers,sort&compress]{natbib}
\usepackage{comment}
\usepackage{longtable}
\usepackage{multirow}
  
\usepackage[bottom]{footmisc}
\usepackage{booktabs}

\usepackage{float}
\usepackage{subfig}
\usepackage{wrapfig}
\usepackage{framed,xcolor}
\usepackage{newfloat}
\usepackage[xcolor,framemethod=TikZ]{mdframed}
\usepackage{mathpazo}
\usepackage{amsthm}
\usepackage{thmtools}
\usepackage{authblk}
\patchcmd{\thebibliography}{\section*{\refname}}{}{}{}
\usepackage{widetext}
\usepackage{tikz}
\usetikzlibrary{arrows,shapes,chains,calc,patterns,decorations.pathreplacing}

\newcommand{\GX}{\mathcal{G}}

\newcommand{\NX}{\mathcal{N}}

\newcommand{\MM}{\mathcal{M}}

\newcommand{\relu}{\mathsf{ReLU}}

\newtheoremstyle{theoremdd}
{\topsep}
{\topsep}
{\itshape}
{0pt}
{\fontfamily{cmss}\selectfont\bfseries}
{.}
{ }
{\thmname{#1}\thmnumber{ #2}\thmnote{ (#3)}}

\theoremstyle{theoremdd}

\mdfdefinestyle{myStyle}{roundcorner=5pt,backgroundcolor=brown!20,linecolor=red!40!black,linewidth=1.5pt}

\usepackage{titlesec}
\titleformat*{\section}{\fontfamily{cmss}\selectfont\large\bfseries}
\titleformat*{\subsection}{\fontfamily{cmss}\selectfont\normalsize\bfseries}
\titleformat*{\subsubsection}{\fontfamily{cmss}\selectfont\normalsize}
\usepackage[left=0.75 in, right=0.75 in, top=1.00 in, bottom=1.00 in]{geometry}
\graphicspath{{./Figures/}}

\newcommand\blfootnote[1]{%
	\begingroup
	\renewcommand\thefootnote{}\footnote{#1}%
	\addtocounter{footnote}{-1}%
	\endgroup
}

\begin{document}

		
		
		\title{\fontfamily{cmss}\selectfont Traffic Data Imputation using Deep Convolutional Neural Networks}
		


\author[1]{Ouafa Benkraouda}
\author[1]{Bilal Thonnam Thodi}
\author[3]{Hwasoo Yeo}
\author[4]{Monica Menendez}
\author[1,4]{Saif Eddin Jabari$^{\star,}$}

\affil[1]{University of Illinois at Urbana-Champaign, Champaign, IL 61820, U.S.A.}
\affil[2]{New York University Tandon School of Engineering, Brooklyn NY, U.S.A.}
\affil[3]{Korea Advanced Institute of Science and Technology, Daejeon, Republic of Korea}
\affil[4]{New York University Abu Dhabi, Saadiyat Island, P.O. Box 129188, Abu Dhabi, U.A.E.}

\date{}

\twocolumn[
\begin{@twocolumnfalse}
	
\maketitle	

{ \fontfamily{cmss}\selectfont\large\bfseries		
\begin{abstract}
{ \normalfont\normalsize
	We propose a statistical learning-based traffic speed estimation method that uses sparse vehicle trajectory information. Using a convolutional encoder-decoder based architecture, we show that a well trained neural network can learn spatio-temporal traffic speed dynamics from time-space diagrams. We demonstrate this for a homogeneous road section using simulated vehicle trajectories, and then validate it using real-world data from NGSIM. Our results show that with probe vehicle penetration levels as low as 5\%, the proposed estimation method can provide a sound reconstruction of macroscopic traffic speeds and reproduce realistic shockwave patterns, implying applicability in a variety of traffic conditions. We further discuss the model's reconstruction mechanisms and confirm its ability to differentiate various traffic behaviors such as congested and free flow traffic states, transition dynamics, and shockwave propagation.
	
	\medskip
	
	\textbf{\fontfamily{cmss}\selectfont Keywords}: Convolutional neural networks, data expansion, data imputation, estimation, filtering, traffic dynamics, traffic state estimation.
	
}
\end{abstract}
}
\bigskip
\end{@twocolumnfalse}
]

		
	
	
	

\section{Introduction}
\label{S:intro}
Recent\blfootnote{$^{\star}$ Corresponding author, Email: \url{sej7@nyu.edu}} studies have shown that connected and autonomous vehicles (CAVs) can ease traffic flow instabilities at low CAV penetration levels \citep{bayen2018flow,stern2018stopgo,stern2017singleauto,kaidi2018weavingauto}. To do so, these systems require accurate knowledge of traffic conditions.  In general, most advanced traffic management and control tools require accurate inputs. However, measurements remain sparse in today's road traffic networks and high penetration levels of CAVs are not expected in the coming decade or two. In the meantime, traffic state reconstruction methods are needed to fill this gap. Most traffic state estimation techniques in the literature have focused on reproducing traffic conditions at aggregate levels (averaged over segment lengths greater than 100 meters and time intervals greater than one minute \citep{polson2017deep,koes2016dnn-flow}). Advanced traffic management  tools (e.g., adaptive traffic signal control) require knowledge of traffic conditions at a much finer scale \citep{jabari2018stochastic,zheng2018stochastic,jabari2019sparse,Kaidi2016TState,Guler2014TState}; see \citep{seo2017traffic} for a comprehensive review of past traffic state estimation methods.

We present here a data driven methodology to estimate traffic speeds across a road section at fine spatio-temporal resolutions and using limited probe vehicle information. Characterizing traffic speed across small intervals of space and time is especially challenging because of the inherent stochastic nature of traffic flow and noisy sparse traffic observations \citep{jabari2014stochastic,jabari2012stochastic,kaidi2019queueest}. Researchers often resort to different statistical learning methods to model traffic speed variations from historical datasets \citep{ma2015dnn-speed,yuhan2016dnn-speed,bayen2012bayesnet}. However, the majority of these methods only succeed in capturing recurring traffic patterns such as daily variations or within-day variations, and do not necessarily learn the actual flow dynamics. A few studies have focused on modeling the sharp discontinuities found in spatio-temporal traffic dynamics (e.g., caused by transition from free flow to congested state). Nonetheless, the low interpretability of these non-linear models poses reliability and robustness issues when applied \citep{polson2017deep,wang2016lstm,yuhan2016dl-speed}.

Various kernel-based estimation schemes have also been proposed to interpolate spatio-temporal macroscopic traffic speeds from heterogeneous data sources, mainly, loop detectors and probe vehicles \citep{treiber2002filter,trieber2011filter, chen2019filter,ambhul2016mfd,dakic2018lagrang}. However, these interpolation methods need field calibration of static model parameters (offline or online) such as shockwave speeds and free flow speeds, which can dynamically change depending on the local-traffic conditions, leading to biased results.

In this paper, we propose a method for traffic state reconstruction of traffic speeds dynamics using vehicle trajectory information obtained from sparse probe data (with as low as $5 \%$ penetration rate). We employ a deep convolutional neural network (CNN) \citep{goodfellow2016dl,krizhevsky2012imagenet} to learn the traffic speed dynamics from a time-space diagram plotted using partial observations of vehicle trajectories. Using an encoder-decoder architecture \citep{jon2011autoenc}, we develop a CNN that learns the local spatial and temporal correlations (or features) from the time-space traffic speed profiles, and reconstructs the full (macroscopic) traffic speed dynamics for any given sparse vehicle trajectories. We show that a well trained deep CNN can learn to reproduce short-term non-recurring traffic features such as congestion waves and free flow speed dynamics, which state-of-the-art estimation methods often fail to capture. The proposed learning method can adapt to local traffic speed conditions, and reproduce observed shock-wave patterns and stop-and-go traffic. We further provide critical insights into the learned neural network model to understand the essential features detected while reconstructing the full speed map. Such insights can help unlock some of the black-box features of neural networks when applied to traffic state estimation problems.

In the rest of the paper, we formally state the traffic speed estimation problem and motivate the use of an encoder-decoder based convolutional neural network architecture to learn the spatio-temporal traffic speeds in Sec. \ref{sec:method}. We perform numerical experiments using the proposed model on a homogeneous road section and discuss its efficacy in reconstructing the traffic speed maps on a simulated scenario, as well as on the Next Generation Simulation (NGSIM) dataset in Sec. \ref{sec:casestudy}. This is followed by a brief discussion of the model's interpretability and the role played by the hidden layers in the estimation of speed dynamics. We conclude the paper with a discussion of the main insights and suggest potential future research directions in Sec. \ref{sec:conclusion}.

	

\section{Methodology}
\label{sec:method}

\subsection{Traffic Speed Estimation Problem}
\label{subsec:problem-set}
Consider a set of vehicles $\NX$ passing through a given road section of length $X = |\mathcal{X}|$ over time period $T = |\mathcal{T}|$. Let $(x_i,t_i,v_i)$ denote the coordinates of vehicle $i \in \NX$, where $x_i$ and $v_i$ represent its relative longitudinal position (with respect to the starting position of the road section) and speed at time $t_i$. We denote by $\GX := \{(x_i,t_i,v_i)~|~i \in \NX \}$ the set of all $\NX$ vehicle trajectories. Let $\GX_p \subseteq \GX$ represent the trajectories of sampled vehicles in $\NX$ (with sampling percentage $p$), within the space-time domain $X \times T$. Note that the cardinalities of $\GX$ and $\GX_p$ can vary depending on the number of vehicles in the system and the resolution of vehicle coordinates (the sampling cadence).

\begin{figure}[h!]
	\centering
	\subfloat[][All vehicles]{\resizebox{0.3\textwidth}{!}{
			\includegraphics[width=0.3\textwidth,origin=c]{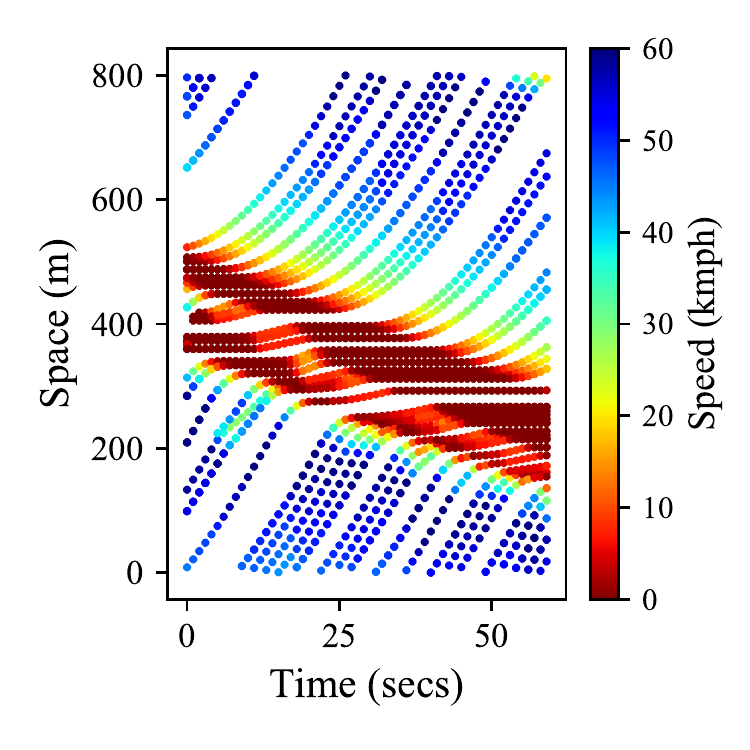}} 
		\label{fig:veh-trja}} 	
		
	\subfloat[][10\% sampled vehicles]{\resizebox{0.3\textwidth}{!}{
			\includegraphics[width=0.3\textwidth,origin=c]{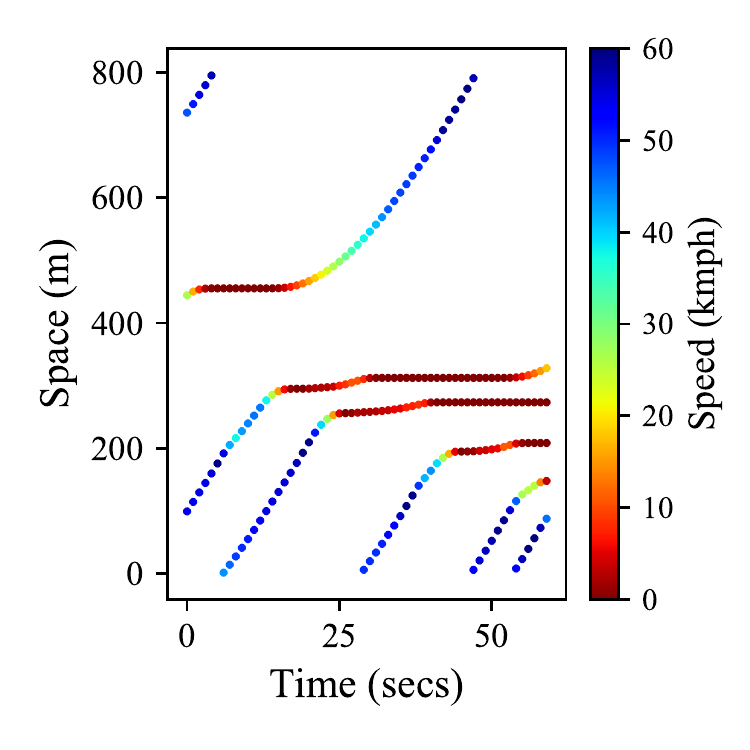}} 
		\label{fig:veh-trjb}}
	\caption{\small Vehicle trajectories plotted on a space-time diagram and color coded with speeds ($X = 800$ m, $T = 60$ sec). We convert these into three-dimensional tensors, which serve as output and input, respectively, to the speed reconstruction model.}
	\label{fig:veh-trj}
\end{figure}

The estimation problem is to determine an arbitrary function from the domain of partial trajectories $\GX_p$ to the full trajectories $\GX$ (or some function of $\GX$, such as macroscopic traffic speeds or traffic density). However, the variable (and not necessarily known) sized nature of $\GX$ (and $\GX_p$) presents a challenge, as a unique mapping function may not exist. Hence, we represent $\GX$ (and $\GX_p$) on a two-dimensional space-time plane as plotted vehicle trajectories, and color code it with the respective speed data; see Fig.~\ref{fig:veh-trj} for an example. 

To interpret this graphical representation, we discretize the entire space-time domain into finite regions called cells, each of which has dimensions $x$ and $t$, the cell's length and time duration, respectively. The values in each cell represent the traffic state in that cell, defined by a three-scale color value based on the RGB format. The three-scale color representation of traffic states allows us to clearly differentiate between the empty cells (where there are no vehicles), free flowing cells (where vehicles travel at the free-flow speed), and heavily congested cells (where vehicles cannot move). For example, empty cells are represented by a white color with RGB value $(255, 255, 255)$ whereas heavily congested cells are represented by a red color with RGB value $(231, 11, 5)$. Colors ranging between red and blue represent traffic states in between free-flow and heavily congested. Using this representation, we encode the vehicle trajectory information as a three-dimensional tensor, denoted by $\mathbf{z} \in  \{1,\dots,255\}^{X \times T \times C}$, where $C$ is the number of color channel arrays. Note here that when treating the time-space diagrams as images using RGB, the tensor values are restricted to the set $\{1,\dots,255\}$. Let $\mathbf{z}^{\mathrm{p}}$ denote the respective tensor with partial trajectory information, and $\textbf{z}^{\mathrm{f}}$ the tensor with full vehicle trajectory information. The traffic state estimation problem can be stated as one that seeks to determine a mapping function $g$, such that $g(\mathbf{z}^{\mathrm{p}})=\mathbf{z}^{\mathrm{f}}$, where $\mathbf{z}^{\mathrm{p}}$ and $\mathbf{z}^{\mathrm{f}}$ are now fixed sized tensors. We refer to the mapping function $g$ as the \emph{traffic speed reconstruction model}.

\subsection{Representation of Speed Reconstruction Model}
\label{subsec:model-form}
\begin{figure}[h!]
	\centering
	\subfloat[][Convolutional Encoder-decoder architecture]{\resizebox{0.49\textwidth}{!}{
			\includegraphics[width=0.49\textwidth,origin=c]{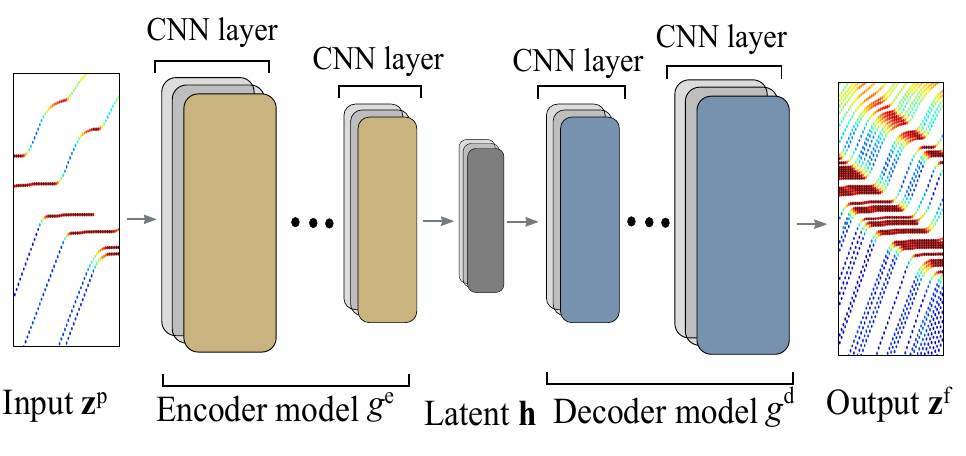}} 
		\label{fig:enc-deca}} 		
	
	\subfloat[][A single CNN layer in the encoder model]{\resizebox{0.49\textwidth}{!}{
			\includegraphics[width=0.49\textwidth,origin=c]{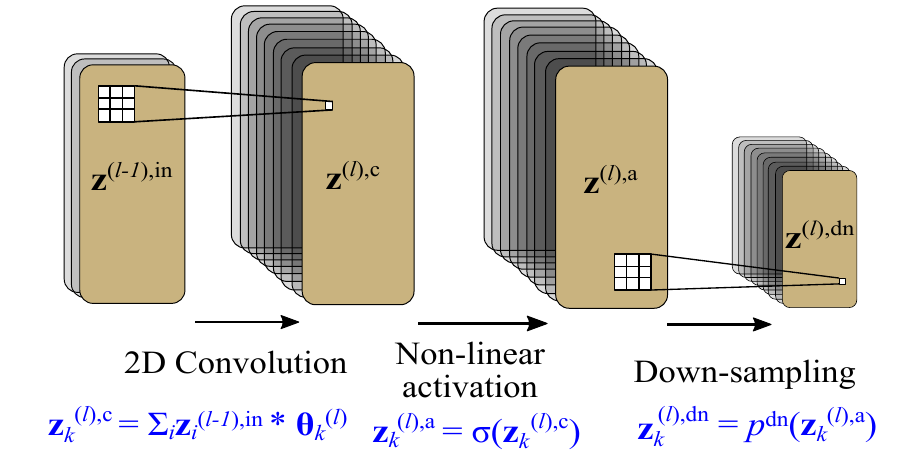}} 
		\label{fig:enc-decb}}
	
		\subfloat[][A single CNN layer in the decoder model]{\resizebox{0.49\textwidth}{!}{
			\includegraphics[width=0.49\textwidth,origin=c]{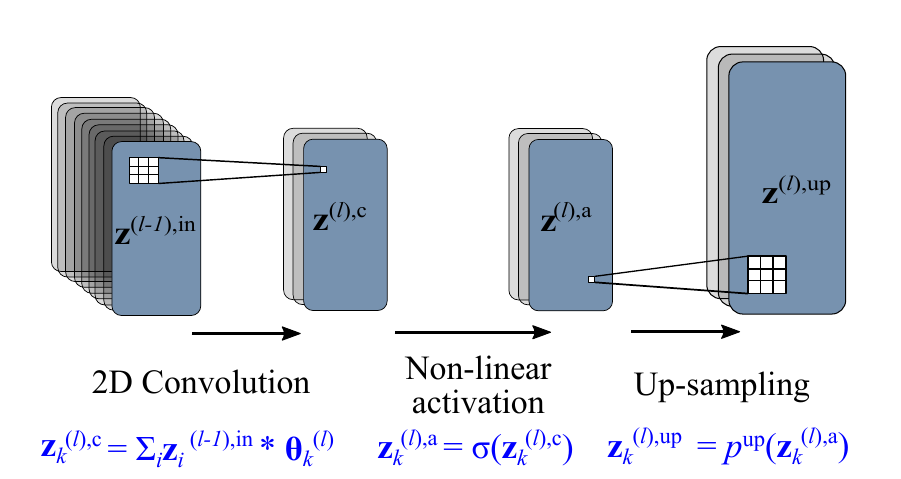}} 
		\label{fig:enc-decc}}
	\caption{\small Representation of the speed reconstruction model using a convolutional encoder-decoder neural network architecture.}
	\label{fig:enc-dec}
\end{figure}

We approximate the speed reconstruction function $g$ using a neural network model. We chose an encoder-decoder convolutional neural network architecture \citep{jon2011autoenc,goodfellow2016dl} to represent the reconstruction model as shown in Fig.~\ref{fig:enc-deca}. The encoder model $g^{\mathrm{e}}$ extracts the primary features from the sparse input vehicle trajectory tensor $\textbf{z}^{\mathrm{p}}$ and maps them into a latent space representation $\textbf{h}$, which can be thought of as an abstract projection of the sparse-vehicle trajectory information contained in $\textbf{z}^{\mathrm{p}}$ onto some higher dimensional space (often unintrepretable). The decoder model, $g^{\mathrm{d}}$, uses the information contained in $\textbf{h}$ to reconstruct the output tensor $\textbf{z}^{\mathrm{f}}$, which in this case is the macroscopic traffic speed map. 

The encoder model ($g^{\mathrm{e}}$) is a sequence of CNN layers $\Lambda$ stacked horizontally, where each layer $l \in \Lambda$ consists of three operations - convolution, non-linear activation and down-sampling (up-sampling for the decoder model); see Fig.~\ref{fig:enc-decb}. The convolution operation is performed using a set of layer-specific kernels 
$\boldsymbol{\Theta}^{(l)} \equiv [\boldsymbol{\Theta}^{(l)}_k]_{k \in \mathcal{K}^{(l)}}$, where $\mathcal{K}^{(l)}$ denotes the set of kernels in layer $l$. Each kernel is a tensor $\boldsymbol{\Theta}^{(l)}_k \in \mathbb{R}^{M \times N \times C}$, where $M \times N$ is the kernel size (a hyper-parameter) and $C$ is the number of color channels used in the feature maps.  The convolution operation is given by
\begin{multline}
\mathbf{z}_{k}^{(l),\mathrm{c}} = \sum_{i \in \mathcal{M}^{(l-1)}} \mathbf{z}_{i}^{(l-1),\mathrm{in}}*\boldsymbol{\Theta}_{k}^{(l)}, \\ k=1,\hdots,K, ~l \in \Lambda,
\end{multline}
where $\mathbf{z}_{i}^{(l-1),\mathrm{in}}$ is the $i$th \emph{input feature map} and $\mathcal{M}^{(l-1)}$ is the set of feature maps in layer $l-1$, while $\mathbf{z}_{k}^{(l),\mathrm{c}}$ is the \emph{convolved feature map} produced using the $k$ kernel in layer $l$. This is followed by a clipping operation using a non-linear activation function $\sigma$:
\begin{equation}
\mathbf{z}_{k}^{(l),\mathrm{a}} =\sigma \big( \mathbf{z}_{k}^{(l),\mathrm{c}} \big), ~ k=1,\hdots,K, ~l \in \Lambda,
\end{equation}
where $\mathbf{z}_{k}^{(l),\mathrm{a}}$ is the $k$th \emph{activated feature map} in layer $l$. The kernels $\boldsymbol{\Theta}^{(l)}$ represent the primary feature detectors in layer $(l)$, where each kernel $\boldsymbol{\Theta}^{(l)}_{k}$ detects distinct patterns (or features) from the input feature maps $\textbf{z}^{(l),\mathrm{in}} \equiv [\textbf{z}_i^{(l)}]_{i \in \MM^{(l)}}$. For instance, the initial layer can detect discrete traffic states such as free flow and congested traffic. Successive layers detect more complex patterns such as  shockwave propagation and transient dynamics. We provide a more detailed interpretation of the kernels in the Results and Discussion section.

In the encoder model (Fig.~\ref{fig:enc-deca}), the activated feature maps undergo a down-sampling operation to produce an abstract (lower dimensional) representation of the feature activations as shown below:
\begin{multline}
\mathbf{z}^{(l),\mathrm{dn}} = p^{\mathrm{dn}} \big( \mathbf{z}^{(l),\mathrm{a}} \big) \\ = \underset{m, n}{\max} \left( \mathbf{z}^{(l),\mathrm{a}} \cdot \mathbf{u}(m ,n) \right), ~ l \in \Lambda,
\label{eqn:maxpool}
\end{multline}
where $\mathbf{u}(m, n)$ is a two-dimensional spatial-window applied to the input tensor $\textbf{z}^{(l),\mathrm{a}}$. Note that these depend on the kernel dimensions $M\times N$.  The down-sampling \eqref{eqn:maxpool} returns the largest value of $\textbf{z}^{(l),\mathrm{a}}$ (a.k.a. max pooling function \citep{scherer2010evaluation}). Note that \eqref{eqn:maxpool} reduces the dimension of $\textbf{z}^{(l),\mathrm{a}}$ by a factor of ($m \times n$), where ($m, n$) represents the max-pooling filter size. The max-pooling operation summarizes the primary features detected over various spatial regions in the time-space diagram using the information contained in its smaller sub-regions.

\begin{figure}[h!]
	\resizebox{0.49\textwidth}{!}{%
		\includegraphics{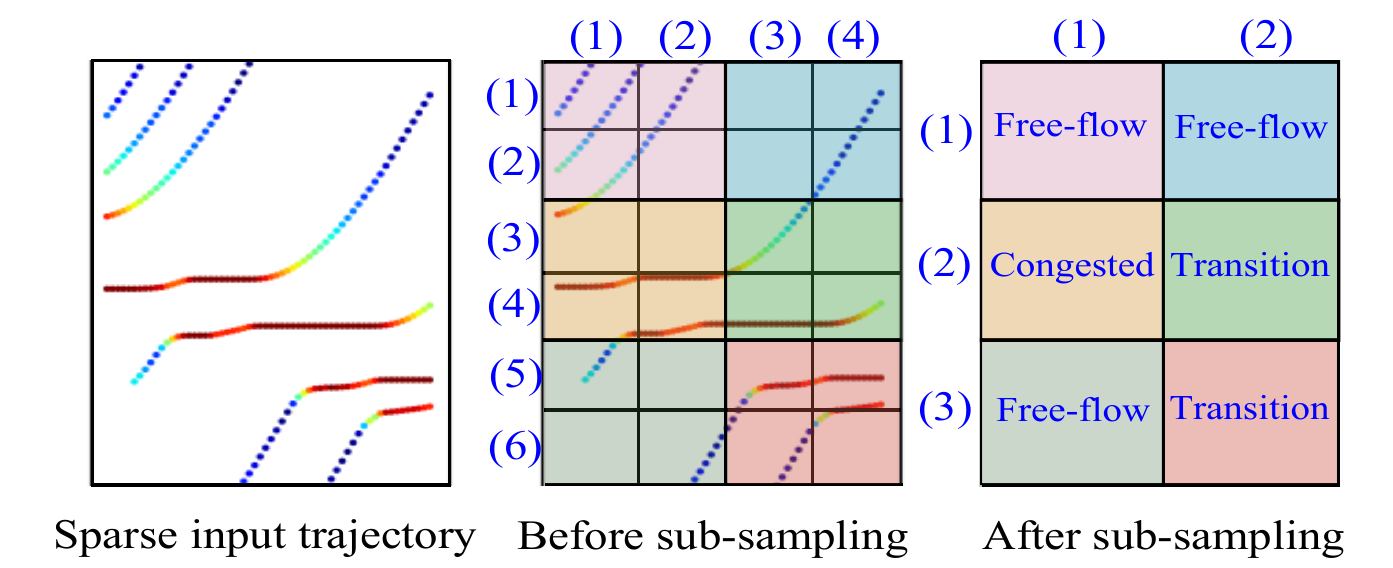}}
	\caption{\small Illustrating the need for down-sampling to handle the sparsity in vehicle trajectory information. The down-sampling operation provides educated guesses of traffic states in each spatial region by inference from traffic states detected in smaller sub-regions.}
	\label{fig:maxpool}
\end{figure}

The down-sampling operation can handle the sparsity in the vehicle trajectory information contained in the input tensor. We illustrate this using an arbitrary (sparse) input vehicle trajectory dataset, and its corresponding activated feature map obtained before and after the down-sampling operation in Fig.~\ref{fig:maxpool}. The traffic state information contained in cells $(2, 3)$ and $(3,  2)$ is absent from the activated feature maps (middle part of Fig.~\ref{fig:maxpool}). However, the down-sampling operation provides an improved estimate of the states of these cells (as free flow and congested states, respectively) by utilizing information contained in neighboring cells (left part of Fig.~\ref{fig:maxpool}). This partially explains why the spatial distribution of probe vehicles along a road section can affect state estimation accuracy \citep{dilip2017sparse,jabari2019learning}.
The output of the down-sampling operation, $\mathbf{z}^{(l),\mathrm{dn}}$ serves as the input into the next layer in the encoder model $g^{\mathrm{e}}$, or to the hidden state representation $\mathbf{h}$. The decoder model $g^{\mathrm{d}}$ also consists of a similar set of layers as $g^{\mathrm{e}}$, except that the down-sampling operation is replaced by an up-sampling operation; see Fig.~\ref{fig:enc-decb}. The most common up-sampling operation is the nearest neighbor function \citep{scherer2010evaluation}, defined as
\begin{equation}
\mathbf{z}^{(l),\mathrm{up}} = p^{\mathrm{up}} \big( \mathbf{z}^{(l),\mathrm{a}} \big) = \mathbf{z}^{(l),\mathrm{a}} \otimes \mathbf{1}_{m^{\prime} \times n^{\prime}}, ~ l \in \Lambda,
\label{eqn:upsam}
\end{equation}
where $\otimes$ is the Kronecker product and $\mathbf{1}_{m^{\prime} \times n^{\prime}}$ is an $m^{\prime} \times n^{\prime}$ matrix of ones.  The dimensions $m^{\prime} \times n^{\prime}$ depend on the size of the full feature maps $X \times T$. The up-sampling operation in \eqref{eqn:upsam}, besides scaling the activated feature maps by ($m^{\prime} \times n^{\prime}$), enforces a plausible and logical reconstruction of traffic state by drawing inferences from super-regions. For example, in left part of Fig.~\ref{fig:maxpool}, one can reconstruct the traffic state in its sub-regions by logically looking at the states in the neighboring regions.

The motivation to use a CNN layer as opposed to a naive feed forward neural network in the encoder-decoder architecture (Fig.~\ref{fig:enc-dec}) is the parameter sharing and sparse connectivity properties of CNNs \citep{goodfellow2016dl}. For traffic data, parameter sharing implies that a specific set of learned features $\boldsymbol{\Theta}^{(l)}$ (such as for recognizing the free flow speed, congested traffic, and transition dynamics) can be used anywhere in the time-space plane (in other words, it is translation invariant). Sparse connectivity ensures that the speed estimates in certain regions in space and time heavily depend on the states in the immediate (local) surroundings, which is especially true for traffic dynamics \citep{trieber2011filter}.  Moreover, the estimation from sparse data problem can be thought of as a \emph{spatial imputation/interpolation problem}, to which CNNs are well suited.

\subsection{Parameter Optimization}
\label{subsec:params-opt}

The output from the decoder model, $g^{\mathrm{d}}$ is the reconstructed speed map denoted as $\hat{\mathbf{z}}_{\boldsymbol{\Theta}}^{\mathrm{f}}$, and is a function of the kernel set $\boldsymbol{\Theta} \equiv \{ \boldsymbol{\Theta}^{(1)}, \dots, \boldsymbol{\Theta}^{(|\Lambda|)} \}$. Thus, the traffic speed estimation problem is cast as a kernel set learning problem:
\begin{equation}
\boldsymbol{\Theta}^* = \underset{\boldsymbol{\Theta}}{\arg \min}~\mathcal{L}(\mathbf{z}^{\mathrm{f}}, \hat{\mathbf{z}}_{\boldsymbol{\Theta}}^{\mathrm{f}}),
\label{eqn:loss-fn}
\end{equation}
where $\mathcal{L}$ denotes a loss function over the estimated and actual speed values. Here, \eqref{eqn:loss-fn} is a differentiable function in $\boldsymbol{\Theta}^{(l)} \in \boldsymbol{\Theta}$, and any gradient based optimization technique can be used to find the optimal kernel set \citep{goodfellow2016dl}. Note that \eqref{eqn:loss-fn} only optimizes the kernel set $\boldsymbol{\Theta}$, and does not include other hyper-parameters such as the number of CNN layers, $|\Lambda|$, in the encoder and decoder model ($g^{\mathrm{e}}$ and $g^{\mathrm{d}}$), the number of kernels in each layer ($\{|\mathcal{K}^{(l)}|\}_{l \in \Lambda}$), the kernel dimensions ($M$ and $N$), the down-sampling filter dimensions ($m$ and $n$), and the up-sampling filter dimensions ($m^{\prime}$ and $n^{\prime}$). These hyper-parameters have to be fine tuned separately.

\section{Experiments and Results}
\label{sec:casestudy}

Here we demonstrate the proposed speed reconstruction model for a homogeneous road segment (a freeway section) using both simulated data and NGSIM data. By looking at different initial and boundary conditions, we produce various traffic conditions corresponding to free-flow, congested, and stop-and-go traffic. The trajectories of all vehicles passing through the segment are recorded at a one second cadence.

\subsection{Learning Speed Reconstruction Model from Data}
\label{model-learn}
Using the simulated vehicle trajectories (for a 3-lane road that is $800$ m long), we generate input-output samples for training the speed reconstruction model, with the following parameters: $X=800$ m, $T=60$ sec, $x=10$ m, and $\tau=1$ sec. The input samples are generated using a $p=5\%$ penetration rate sampled from an unknown probability distribution. This ensures that the sampling replicates actual field observations and no bias in the spatial-distribution of probe vehicles is introduced. The output samples are generated using all vehicle trajectories. In each sample, the vehicle trajectories are color-coded using a fixed speed gradient of 0-60 kmph, where $0$kmph is represented by red and $60$ kmph is represented by blue; see Fig.~\ref{fig:veh-trj} for example. These figures are then converted to $80 \times 60 \times 3$ tensor, where $3$ corresponds to the RGB channel arrays.

We now test our model using two different datasets. The first set is generated from the simulated vehicle trajectories for the same road section used to train the model but under different traffic demand conditions. The second set is produced from the NGSIM vehicle trajectory dataset for I-80 \citep{ngsim}. This allows us not only to test the transferability of our model, but to validate it against real world field data, and check its applicability to traffic behaviors not seen in the training sets. For both datasets we sample 5\% of vehicle trajectories as input data.

The encoder and decoder models used here consists of three CNN layers, and the kernel size, number of kernels, and sampling stride for each layer are shown in Table~\ref{tab:model-arch}. We use a random search technique to determine the best number of layers, the number of kernels in each layer, and the kernel sizes. One can also resort to more sophisticated approaches such as Bayesian-optimization to determine the optimal hyper-parameters \citep{Snoek2012hyper}. For all the layers we use a rectified linear unit ($\relu$) for activation (defined as $\relu(z) = \max(z,0)$ for any given input $z$) except for the last layer which has a Sigmoid activation function (defined as $\sigma(z) = (1+\exp(-z))^{-1}$ for any given input $z$). As we treat the reconstructed speed maps as images, the (normalized) model outputs are bounded between 0 and 1, and this can be obtained using a sigmoid activation function.

\begin{table*}[h!]
	\caption{Hyper-parameters for the speed reconstruction model.}
	\label{tab:model-arch}
		\begin{tabular}{@{}lllll@{}}
			\toprule
			\multicolumn{2}{l}{\multirow{2}{*}{\textbf{Layers}}} & \multirow{2}{*}{\textbf{Convolution operation}} & \multirow{2}{*}{\textbf{Non-linearity}} & \multirow{2}{*}{\textbf{Sampling operation}} \\
			\multicolumn{2}{l}{} &  &  &  \\ \midrule
			Input layer &   & Identity & - & - \\ \cmidrule(l){2-5} 
			\multirow{3}{*}{Encoder model} & CNN-1 & Convolution ($5 \times 5 \times 8$) & Relu & Max-pooling ($2 \times 3$) \\
			& CNN-2 & Convolution ($5 \times 5 \times 32$) & $\relu$ & Max-pooling ($2 \times 2$) \\
			& CNN-3 & Convolution ($5 \times 5 \times 64$) & $\relu$ & Max-pooling ($2 \times 2$) \\ \cmidrule(l){2-5} 
			\multirow{3}{*}{Decoder model} & CNN-4 & Convolution ($3 \times 3 \times 64$) & $\relu$ & Nearest neighbour ($2 \times 2$) \\
			& CNN-5 & Convolution ($5 \times 5 \times 32$) & $\relu$ & Nearest neighbour ($2 \times 2$) \\
			& CNN-6 & Convolution ($5 \times 5 \times 8$) & $\relu$ & Nearest neighbour ($2 \times 3$) \\
			Output layer &  & Convolution ($5 \times 5 \times 3$) & $\sigma$ & - \\ \bottomrule
		\end{tabular}
\end{table*}

We use a larger kernel size $(5 \times 5)$ for the initial and the final CNN layers as opposed to the smaller $(3 \times 3)$ kernels commonly found in traditional image processing architectures (see \citep{krizhevsky2012imagenet} for instance). This is because our input tensor is very sparse (only 5-10\% of the pixels have values other than 1), and as the convolution operation primarily detects local correlations in the data, smaller kernels may fail to detect essential features in the input tensor, can hence produce unrealistic results. We overcome this by resorting to larger kernels, which when convolved can better learn the spatio-temporal speed dynamics, especially the backward propagating shockwave patterns, even from sparsely observed speeds.

Notice that the number of kernels in the middle layers of the CNN (the \emph{bottleneck region}) is larger than the number of kernels in the boundary layers of the CNN (see Table~\ref{tab:model-arch}). This is attributed to the feature learning capability of each layer in the network. For instance, kernels in the initial layers can be considered as primary feature detectors, useful for example, to detect colors (corresponding to congested and free-flowing traffic) and slow moving traffic, which are primitive and can be characterized using fewer kernels. However, subsequent layers use these primary features to capture more complex patterns such as free flowing regimes and transition dynamics, which may require a larger number of kernels to capture all relevant patterns. We use an iterative optimization algorithm known as \textbf{Adam} \citep{kingma2014adam} to train the model. Adam optimization is based on first-order gradient descent with an adaptive learning rate, and is found to have good convergence rates in practice. The test results and discussion are presented below.

\subsection{Results and Discussion}

We first provide a visual comparison between the actual vehicle trajectories and the reconstructed speed map for the test datasets, and quantify the reconstruction error using a Structural Similarity Index. We then analyze and interpret the model's reconstruction behavior by visualizing what the kernels learned from the data and its representation behavior in the latent space.

\subsubsection{Reconstruction Performance}

\begin{figure*}[h!]
	\centering
	\resizebox{0.75\textwidth}{!}{%
		\includegraphics{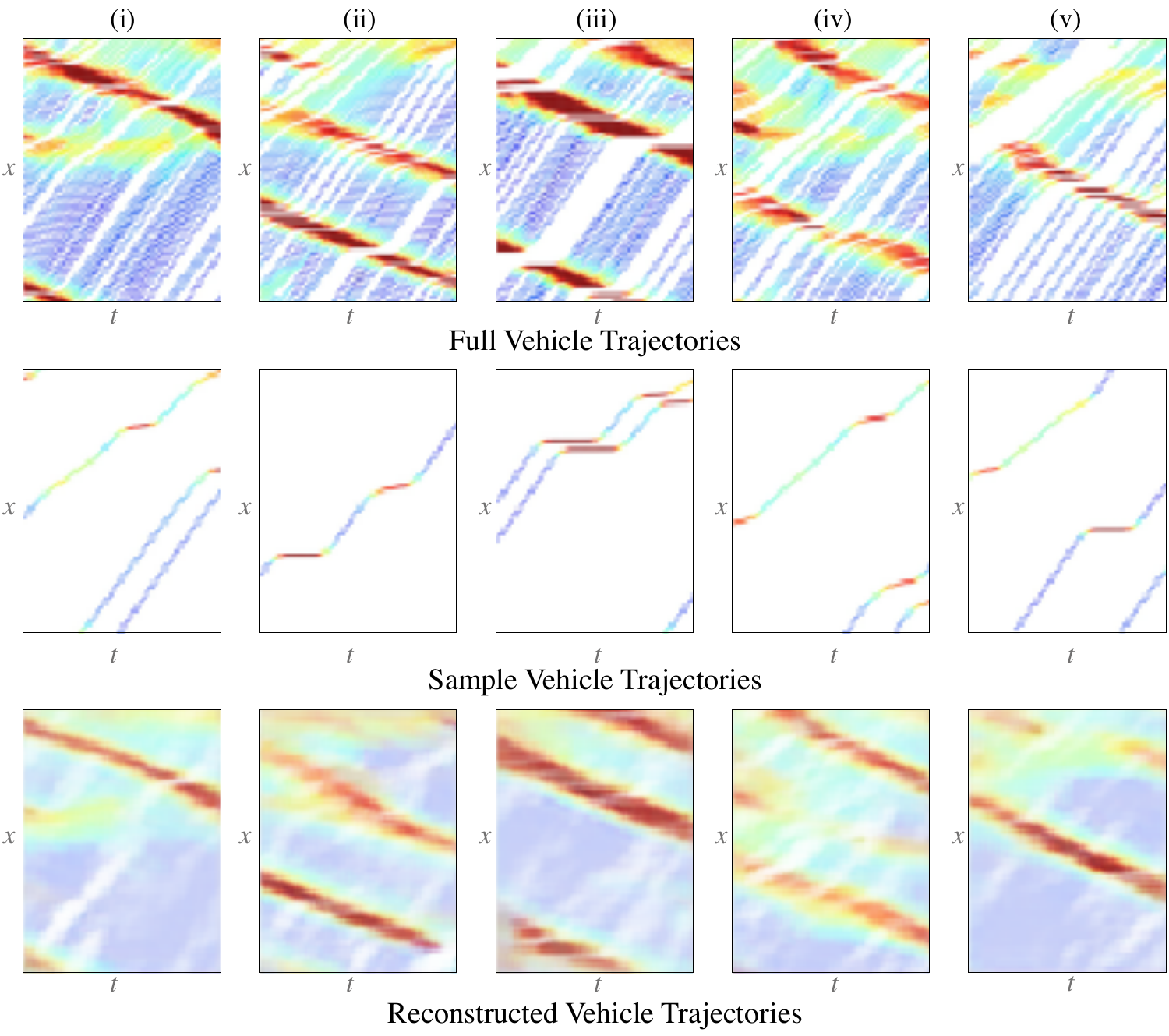}}
	\caption{\small Selected examples where the model accurately reconstructed the macroscopic speed states (simulation data).}
	\label{fig:resultsa}
\end{figure*}

\begin{figure*}[h!]
	\centering
	\resizebox{0.75\textwidth}{!}{%
		\includegraphics{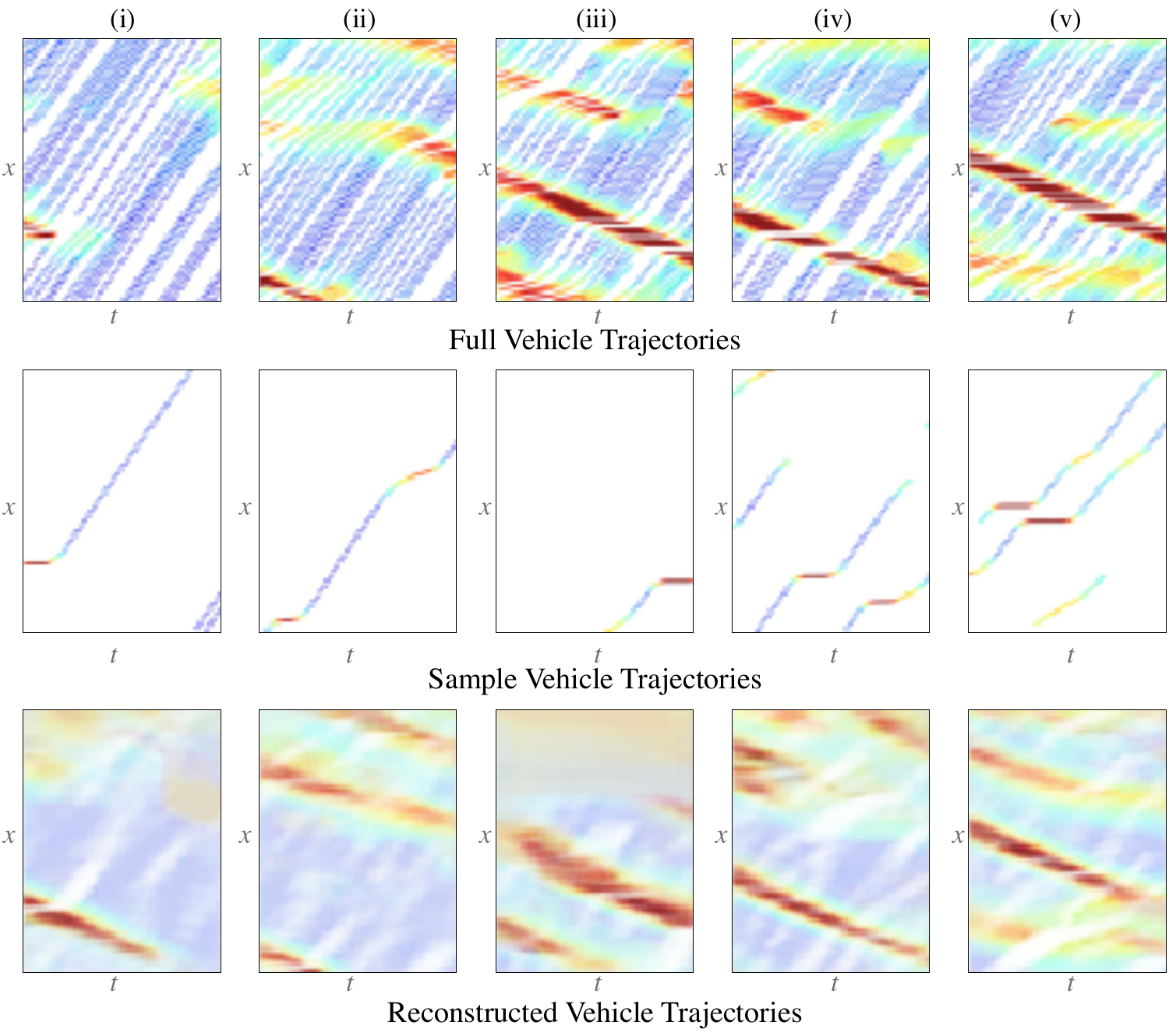}}
	\caption{\small Selected examples where the model partially failed to provide a sound reconstruction of traffic states (simulation data).}
	\label{fig:resultsb}
\end{figure*}

\begin{figure*}[h!]
	\centering
	\resizebox{0.75\textwidth}{!}{%
		\includegraphics{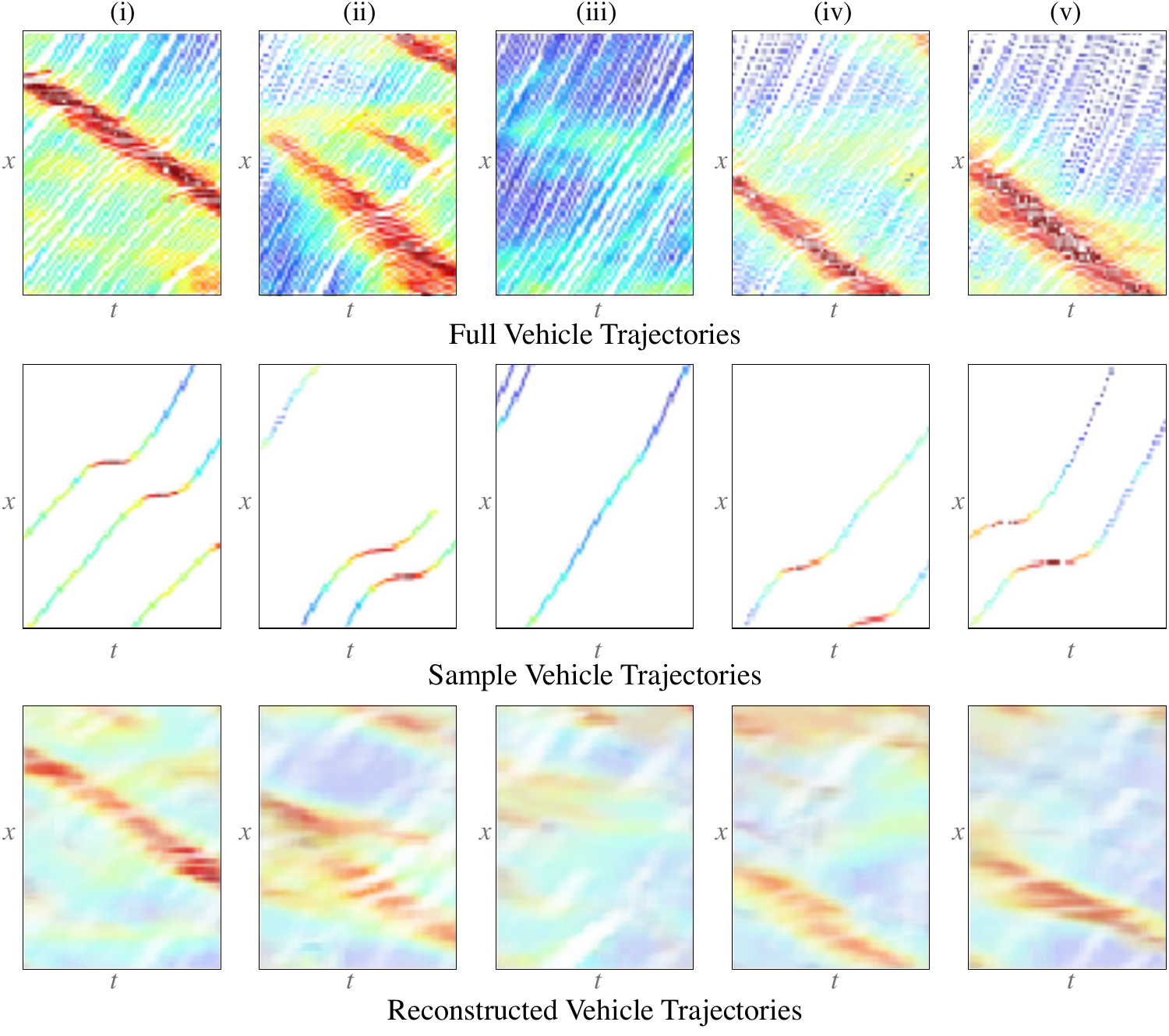}}
	\caption{\small Selected examples from the reconstruction of the NSGIM dataset.}
	\label{fig:resultsc}
\end{figure*}

The test results are presented in Figures \ref{fig:resultsa}-- \ref{fig:resultsc}. From Fig.~\ref{fig:resultsa}, we can see examples where the proposed approach reproduced the traffic speeds correctly. We observe that the model has learned to reconstruct the actual spatio-temporal speed maps and can clearly differentiate between the direction and regime of free flow (blue), congested (red), and transient dynamics (orange, yellow, and cyan).  Interestingly, it succeeds in tracing  backward propagating shockwave patterns by tying together the limited information obtained from the sparse trajectories. In other words, the model is able to reconstruct shockwave patterns with varying sizes and shapes depending on the local traffic conditions, rather than using a mere interpolation assuming a constant wave speed \citep{kaidi2018weavingauto, Xuegang2009Constant,Xuegang2011Constant,Hao2012Constant,Hao2014Constant,Hao2015Constant,Izadpanah2009Constant,Yang2012Constant}.  Fig.~\ref{fig:resultsa} (i, iv, v) shows a scenario where the model captured the dynamic stop-and-go traffic regime (red regions), which cannot be characterized by a single shockwave speed.

Moreover, the model cleverly extrapolates the shockwave occurrence using few speed measurements obtained from the transient dynamics (or the onset of congestion). This can be clearly seen in the lower pane of Fig.~\ref{fig:resultsa} (i, iii), where the probe trajectories only capture slowing traffic  (cyan color), yet the model still managed to detect the occurrence of a shockwave and reconstruct it accordingly. Also, in cases when there was a sufficient number of  representative trajectories, the neural network learned to reconstruct smaller shockwaves and their dissipation behavior as seen from Fig.~\ref{fig:resultsa} (i, iv, v). 

There are, however, a few cases where the method did not perform well. Where probe trajectories are very sparse, the method produces poor estimates of the propagation length and dissipation behavior of the shockwaves. As a result, the reconstructed upstream and downstream jams are either exaggerated or missed; see Fig.~\ref{fig:resultsb} (i, ii, v). Another shortcoming is the model's extrapolation of nonexistent shockwaves. The model learns that with the occurrence of any lower speeds (cyan or yellow), a congestion instance will take place, which is not always true (see Fig.~\ref{fig:resultsb} (i, iv)). This is caused by insufficient information about the local traffic conditions. The backward propagation of stop-and-go traffic depends on the local densities, and if the observed trajectories failed to capture this, then the model may overestimate a shockwave occurrence. Also, if the sample trajectories are very sparse (as in Fig.~\ref{fig:resultsb} (iii)), the model only succeeds in predicting the represented parts and fails elsewhere by completely ignoring the section for which no information exists. Additionally, one can notice the random reconstruction of white areas which depict the absence of vehicle trajectories. If the areas are too large, the model almost always fails in reconstructing these areas as they don't follow any specific pattern. Notice, however, that the undesirable cases described above are \emph{very} few, and not representative of the model's overall performance. They are just included here for completeness.

In order to assess the model's estimation precision with respect to real traffic behavior, we now validate it against NGSIM data Fig.~\ref{fig:resultsc}. The reason behind cross validating our model with unseen and real data is to ensure that the encoder-decoder neural network has in fact learned to mimic traffic behavior rather than memorizing images similar to the training data. Results show that the method successfully captures most of the shockwaves and their varying characteristics. However, since the neural network was only trained on simulation data, it preforms less accurately when estimating shockwaves than in Fig.~\ref{fig:resultsa}. In real traffic, congestion and free flow patterns depend on unobservable aspects that a simulator cannot always reproduce, and therefore we can see how the model didn't accurately reconstruct the shockwave propagation, for example in Fig.~\ref{fig:resultsc} (v). The same applies to the free flow regimes. The method was not trained on cases with pure free flow behavior, which explains the encoder-decoder's poor estimation power in uncongested areas Fig.~\ref{fig:resultsc} (iii, iv). 

\begin{figure}[h!]
	\centering
	\subfloat[][Simulation data]{\resizebox{0.49\textwidth}{!}{
			\includegraphics[width=0.49\textwidth,origin=c]{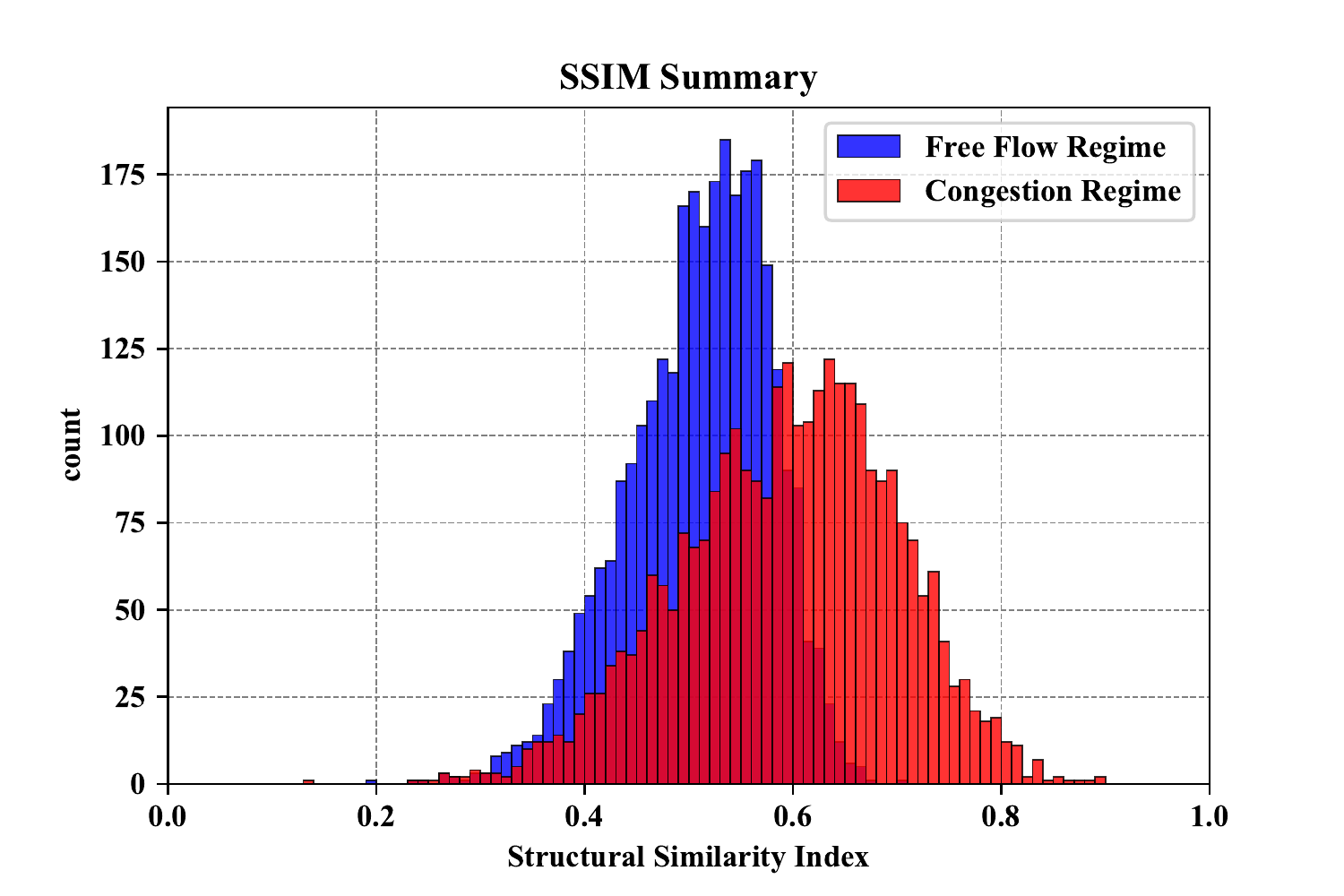}} 
		\label{fig:SSIMa}} 	
	
	\subfloat[][NGSIM data]{\resizebox{0.49\textwidth}{!}{
			\includegraphics[width=0.49\textwidth,origin=c]{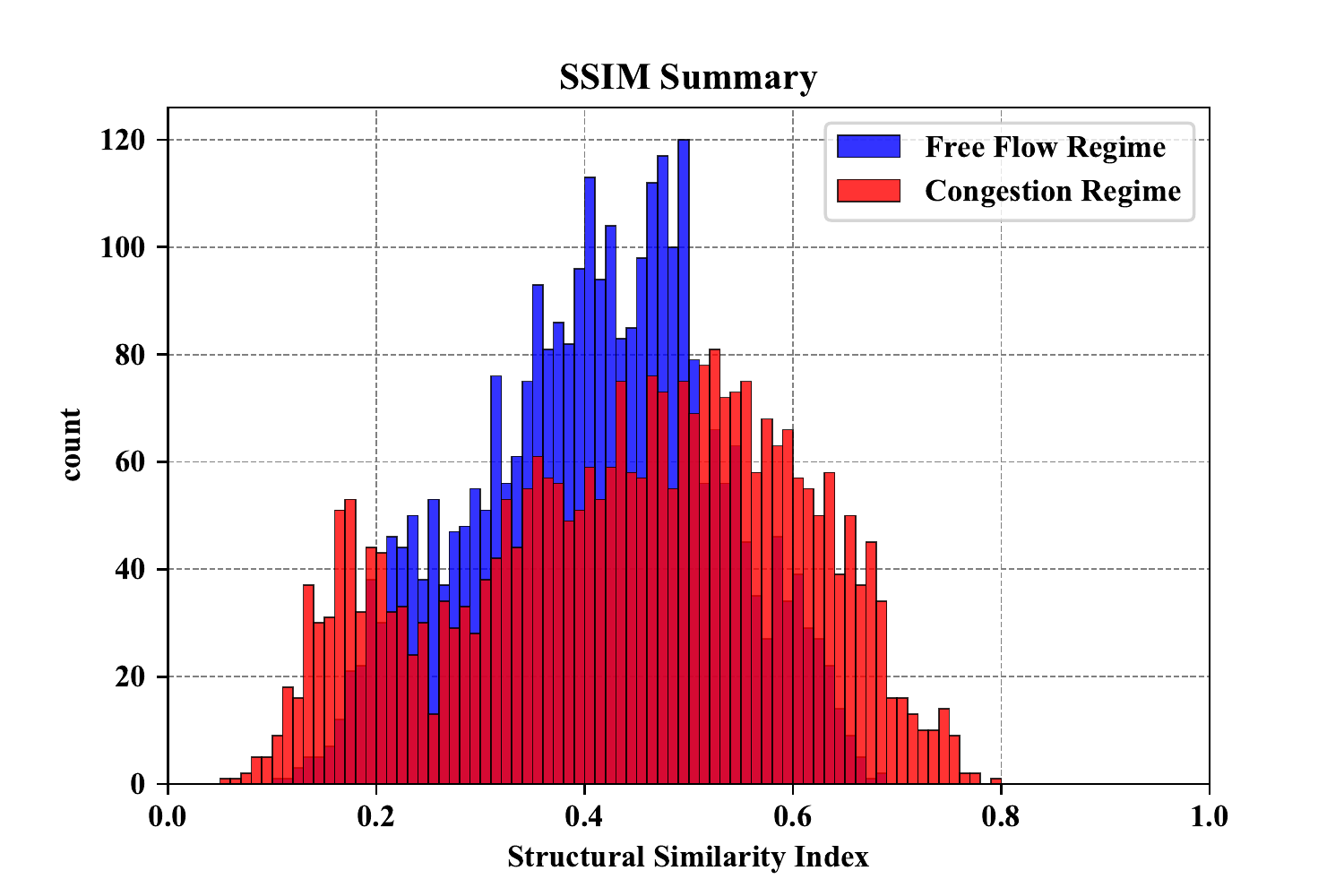}} 
		\label{fig:SSIMb}}
	\caption{\small Histogram showing similarity indices between the actual and reconstructed spatiotemporal maps for the congested and free flow regimes.}
	\label{fig:SSIM}
\end{figure}

To properly quantify the error difference between the reconstructed and the actual traffic speed maps, we use the Structural Similarity Index Method \citep{SSIM}, $\mathsf{SSIM}$. Unlike conventional distance functions such as mean squared error, $\mathsf{SSIM}$ captures changes occurring in the image's structural information. It is defined as:
\begin{equation}
\mathsf{SSIM}(\mathbf{a},\mathbf{b})=\frac{(2\mu_\mathbf{a}\mu_\mathbf{b}+c_1)(2\sigma_{\mathbf{ab}}+c_2)}{(\mu_\mathbf{a}^{2}+\mu_\mathbf{b}^{2}+c_1)(\sigma_\mathbf{a}^{2}+\sigma_\mathbf{b}^{2}+c_2)},
\label{eqn:SSIM-fn}
\end{equation}
where $\mathbf{a}$ and $\mathbf{b}$ represent a sliding Gaussian window with a size of 11x11, $\mu_{\mathbf{a}}$ and $\mu_{\mathbf{b}}$ are the means of $\mathbf{a}$ and $\mathbf{b}$  respectively, $\sigma_{\mathbf{a}}^2$ and $\sigma_{\mathbf{b}}^2$ are the variances of $\mathbf{a}$ and $\mathbf{b}$, respectively, $\sigma_{\mathbf{ab}}$ is the covariance of $\mathbf{a}$ and $\mathbf{b}$, $c_1$ and $c_2$ are stabilizing variables where $c_1 = (k_{1}L)^2$ and $c_2 = (k_{2}L)^2$, $L$ is the dynamic range of pixels, $k_1$ and $k_2$ are default constants set to $0.01$ and $0.03$, respectively (as defined by \citep{SSIM}). The resulting $\mathsf{SSIM}$ index value ranges between $[-1,1]$ where 1 and -1 indicate perfect similarity and 0 indicates no structural similarity. We compute the $\mathsf{SSIM}$ for the free flow and congested regimes separately and the results are shown in Fig.~\ref{fig:SSIM} for the validation dataset.

We observe that the average $\mathsf{SSIM}$ for simulation data in the congested and free flow regimes are $0.6$ and $0.5$ respectively (Fig.~\ref{fig:SSIMa}), implying the model's strong reconstruction power. On the other hand, the average $\mathsf{SSIM}$ values for the NGSIM data are lower with 0.4 and 0.45 for congestion and free flow regimes respectively (Fig.~\ref{fig:SSIMb}). However, having $\mathsf{SSIM}_{\mathrm{cong}} > \mathsf{SSIM}_{\mathrm{free}}$ in both cases, indicates that the model is better at estimating slow-moving traffic than free flowing traffic. This, however, could probably be changed if the model was better trained with pure free flow behavior.

\subsubsection{Interpreting the Speed Reconstruction Model}
We provide here a visual interpretation of the features or patterns that each kernel in the encoder model has learned to detect. Each kernel is activated if it sees a specific pattern in the input space, and this is determined as follows \citep{simonyan2013deepvisual}:
\begin{align}
\mathbf{z}_{k} = \underset{\mathbf{z}_{k}^{(l),\mathrm{a}}}{\arg \max} ~ \mathcal{L}^{\prime} \Big( \mathbf{z}_{k}^{(l),\mathrm{a}} \big( \boldsymbol{\Theta}_{k}^{(l)} \big) \Big),
\label{eqn:kernel-sgd}
\end{align}
where $\mathbf{z}_{k}^{(l),\mathrm{a}} \big( \boldsymbol{\Theta}_{k}^{(l)} \big)$ is the activated feature map in layer $(l)$ using kernel $k$, and $\mathcal{L}^{\prime}(\cdot)$ can be any monotonically increasing function (such as $L^1$-norm or $L^2$-norm) defined over the activated feature map for the kernel $k$ in layer $(l)$. Thus, \eqref{eqn:kernel-sgd} computes an input space $\mathbf{z}_{k}$, which produces the maximum activations for the kernel $\boldsymbol{\Theta}_{k}^{(l)}$. This is shown below in Fig.~\ref{fig:kernel-rep} for a selected subset of kernels in the first 3 layers of our trained speed reconstruction model.
\begin{figure*}[h!]
	\centering
	\subfloat[][Learned kernels in layer 1]{\resizebox{0.95\textwidth}{!}{
			\includegraphics{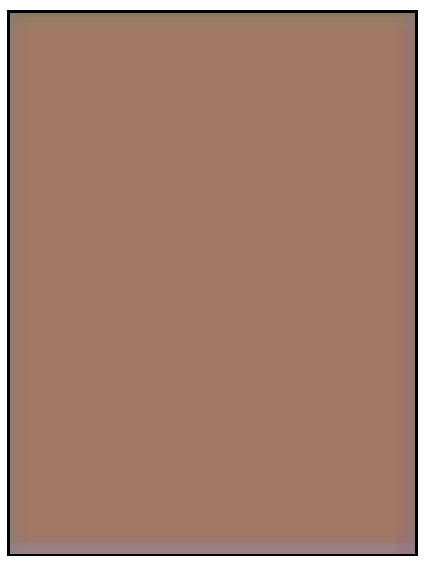}
			\includegraphics{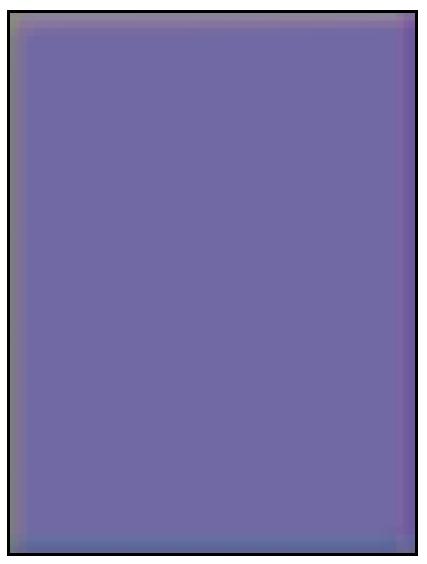}
			\includegraphics{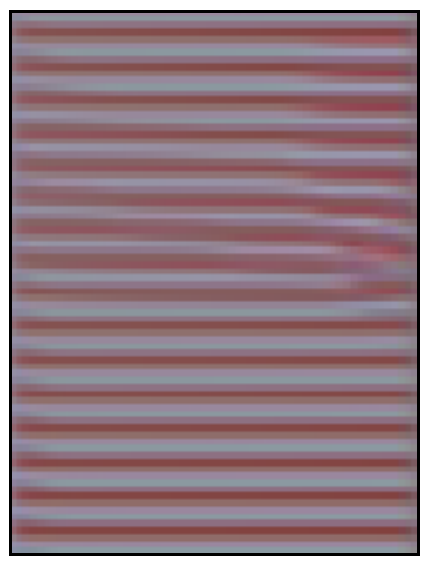}
			\includegraphics{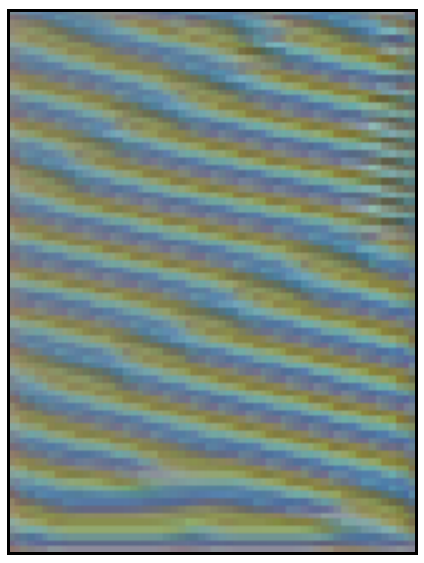}
			\includegraphics{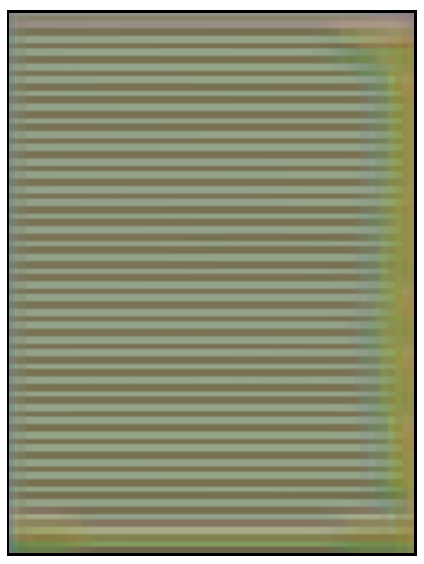}
			\includegraphics{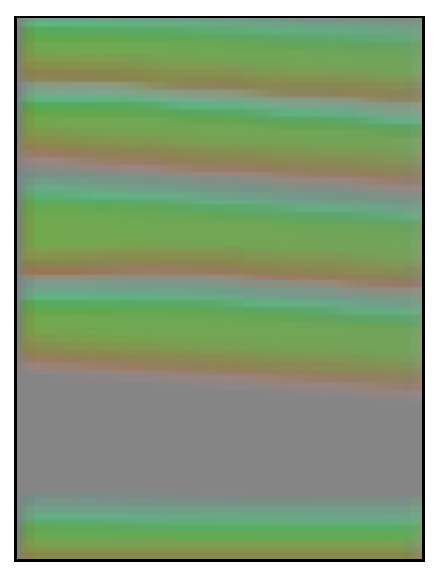}}
		\label{fig:kernel-repa}} 	
	
	\subfloat[][Learned kernels in layer 2]{\resizebox{0.95\textwidth}{!}{
			\includegraphics{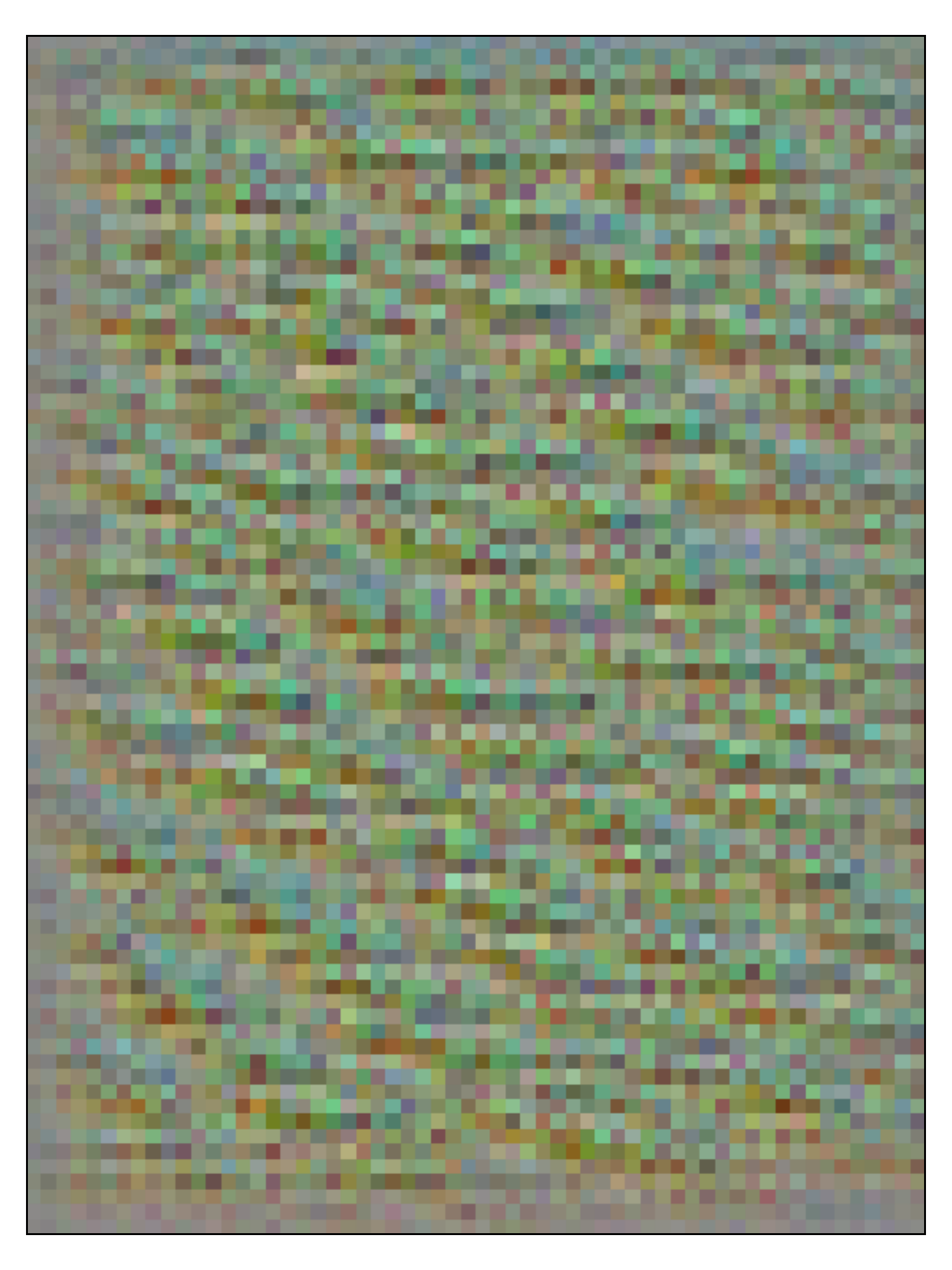}
			\includegraphics{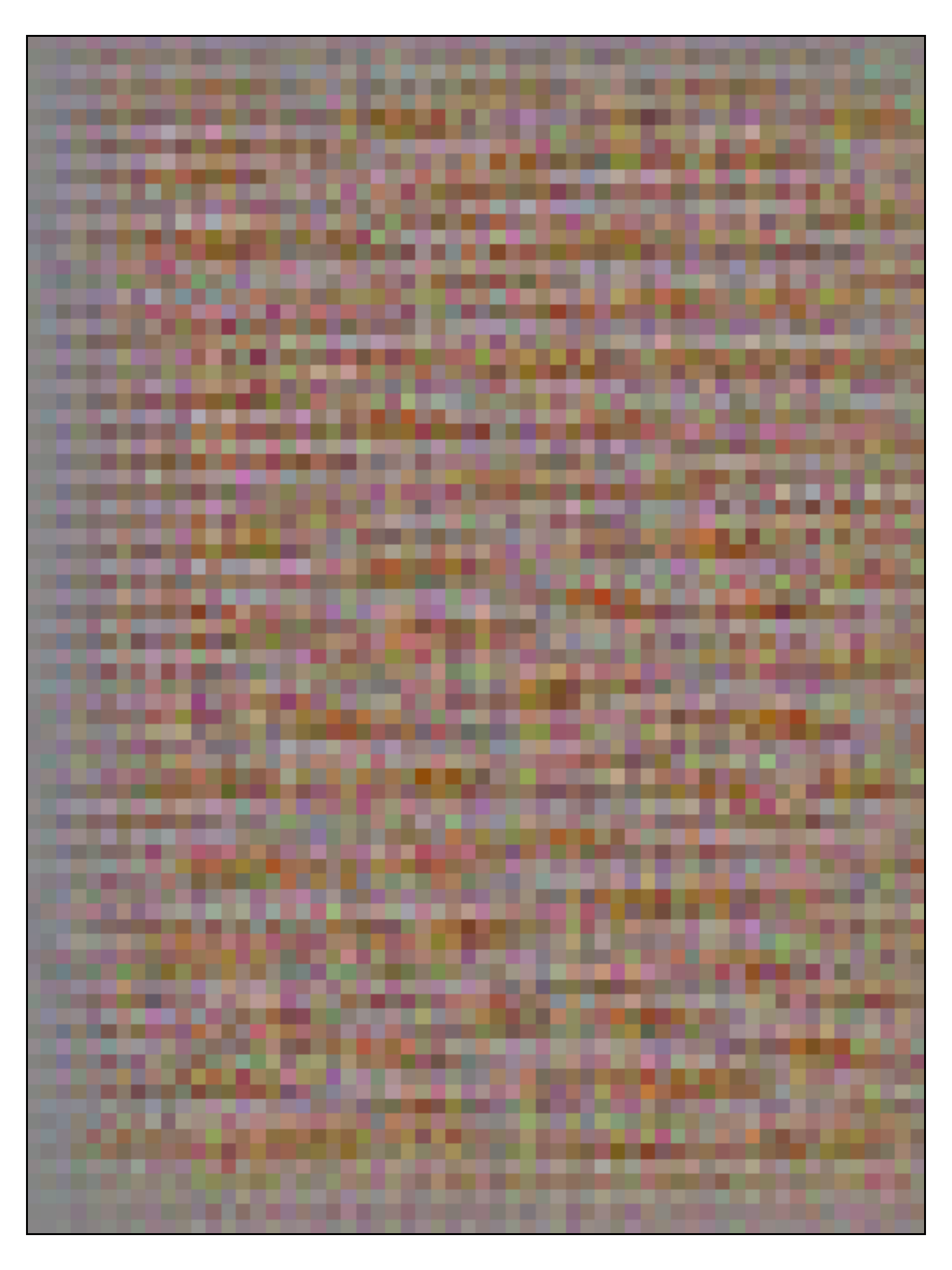}
			\includegraphics{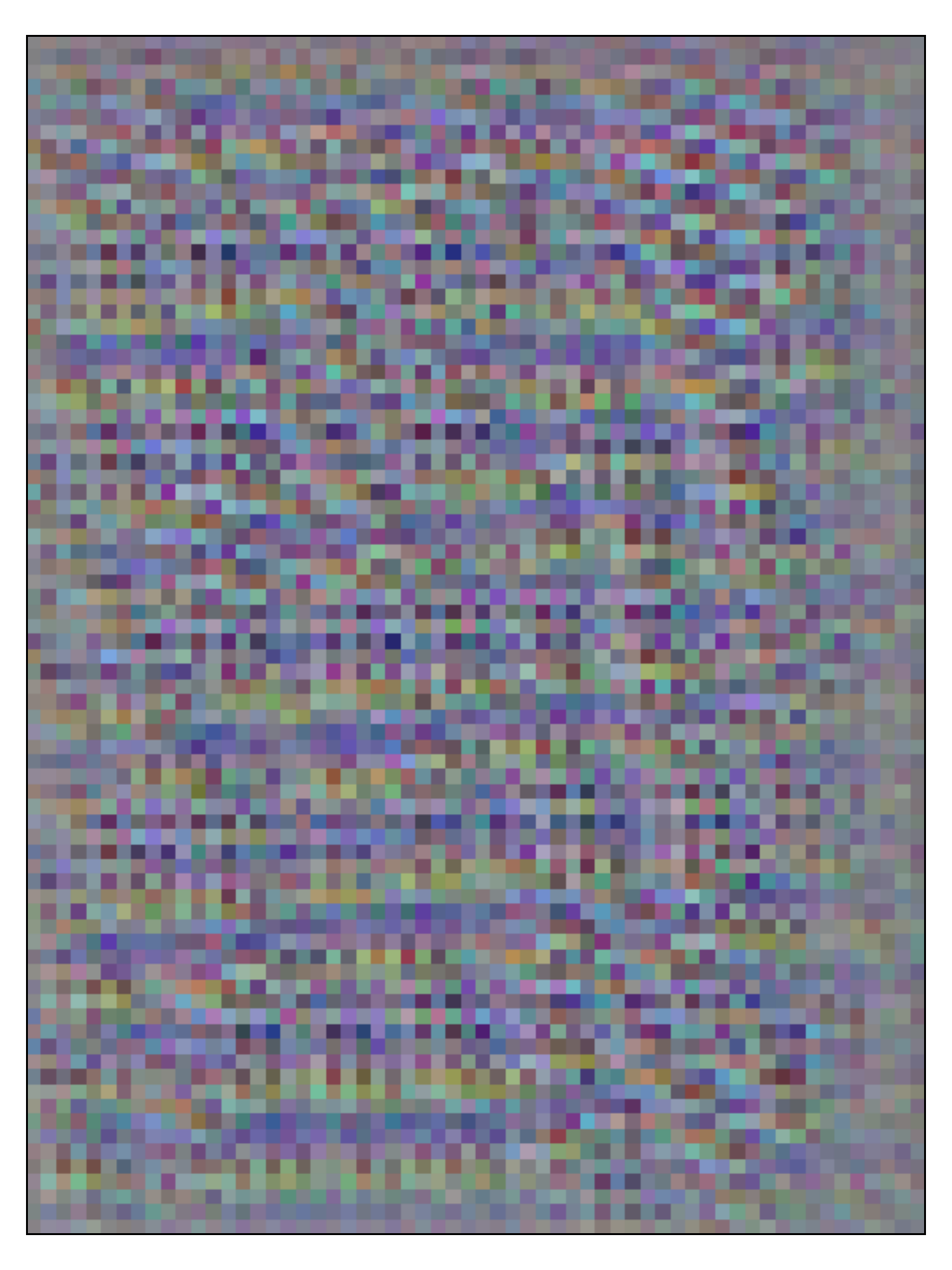}
			\includegraphics{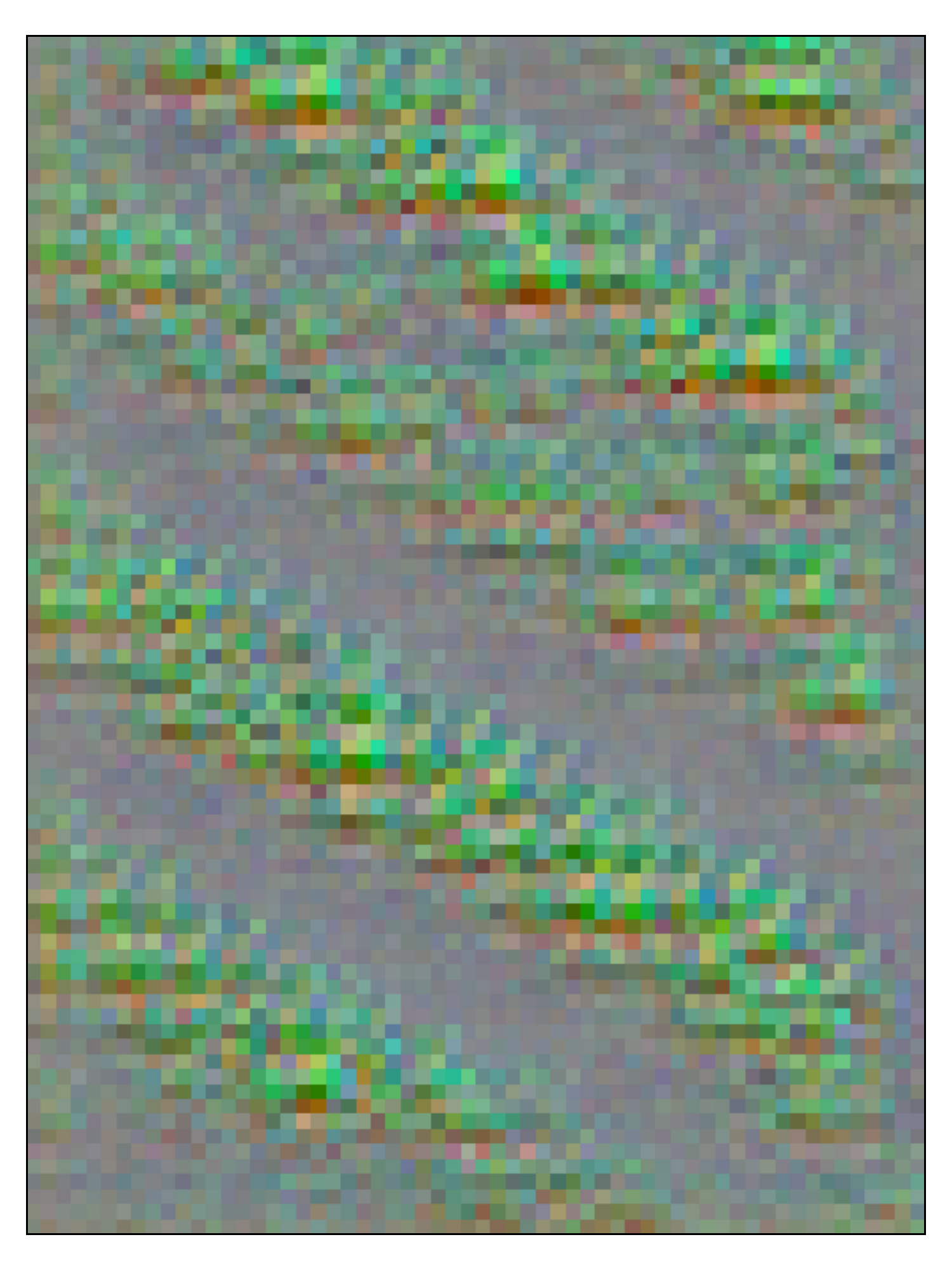}
			\includegraphics{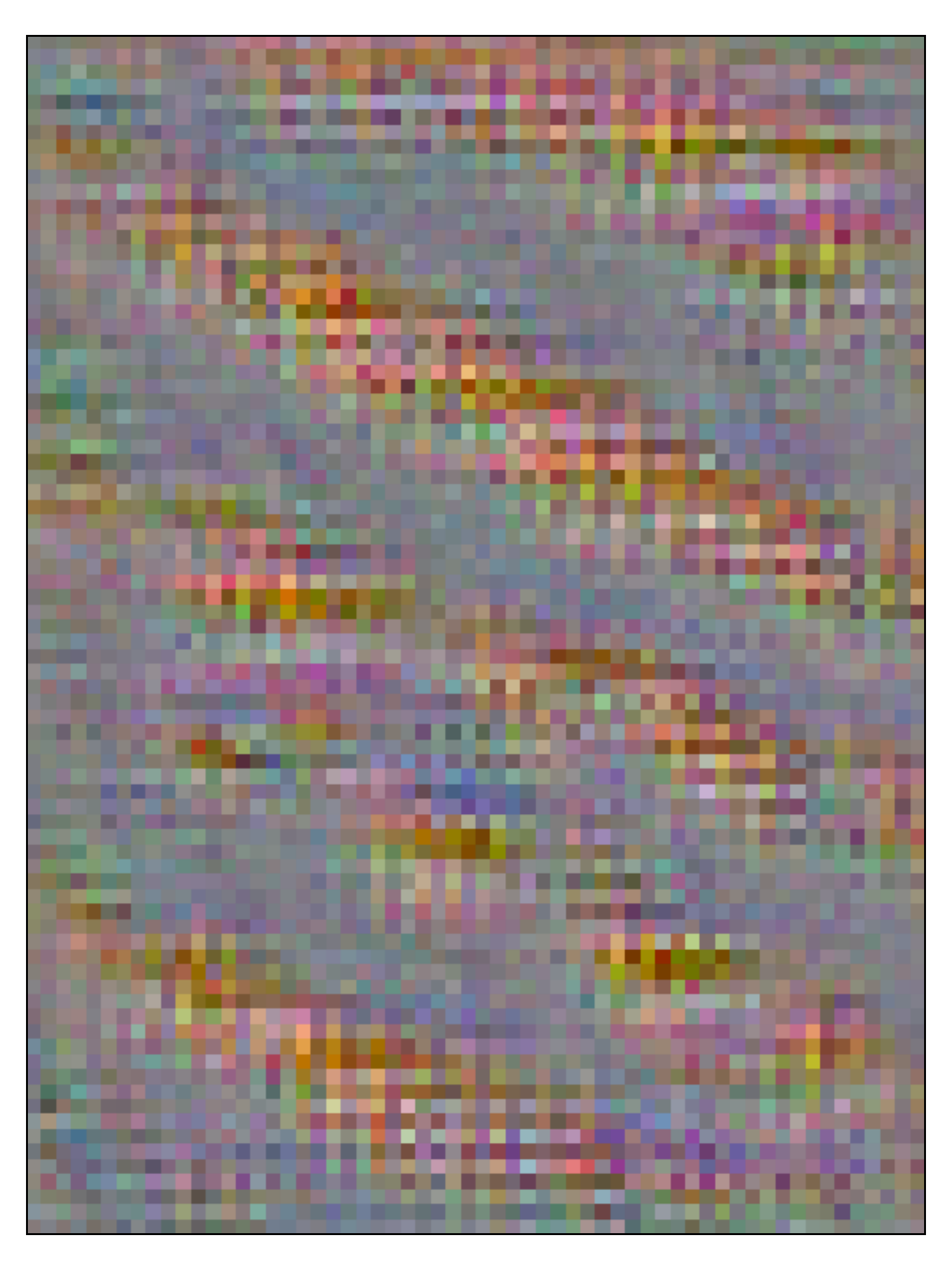}
			\includegraphics{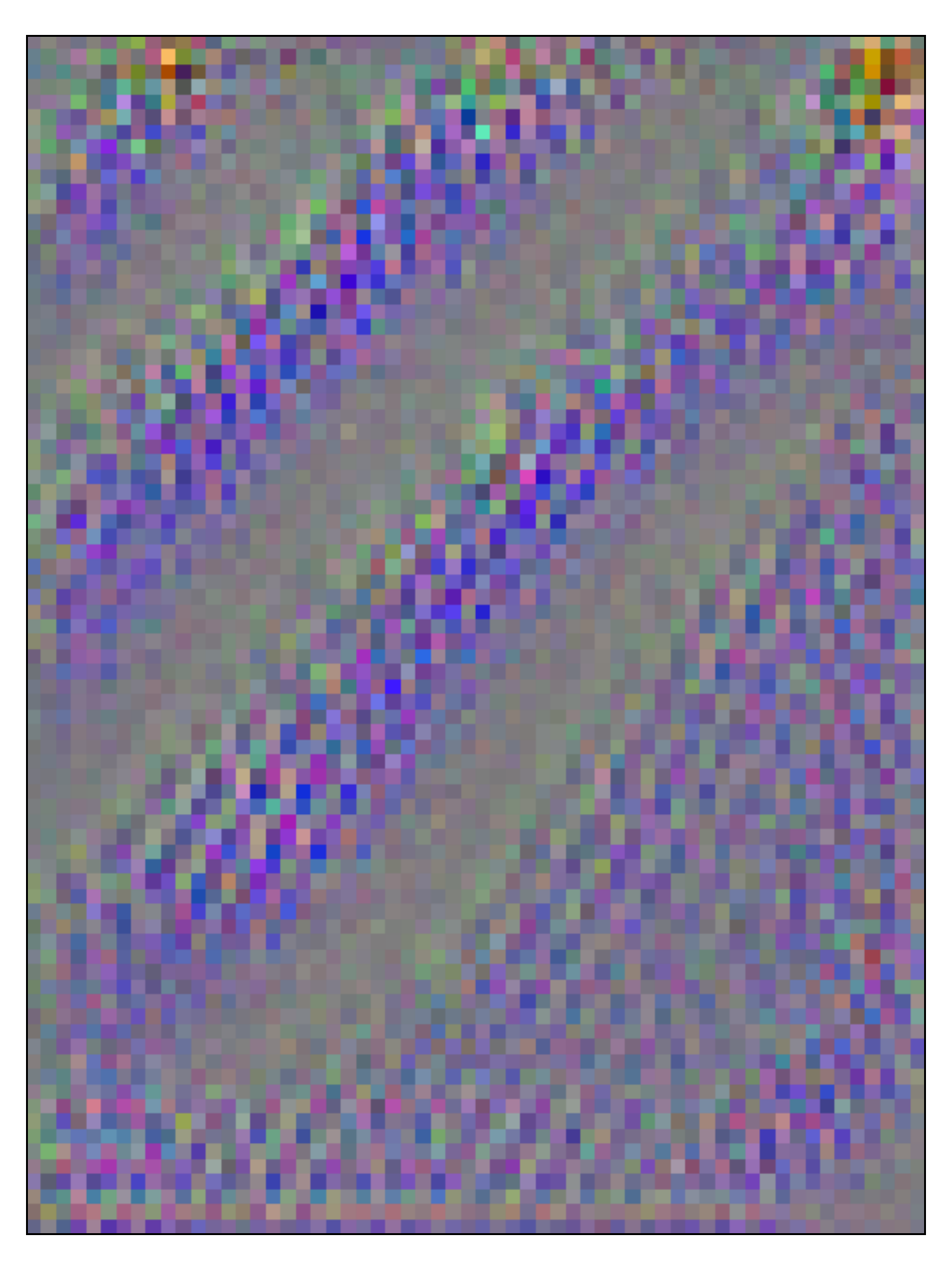}}
		\label{fig:kernel-repb}}
	
		\subfloat[][Learned kernels in layer 3]{\resizebox{0.95\textwidth}{!}{
			\includegraphics{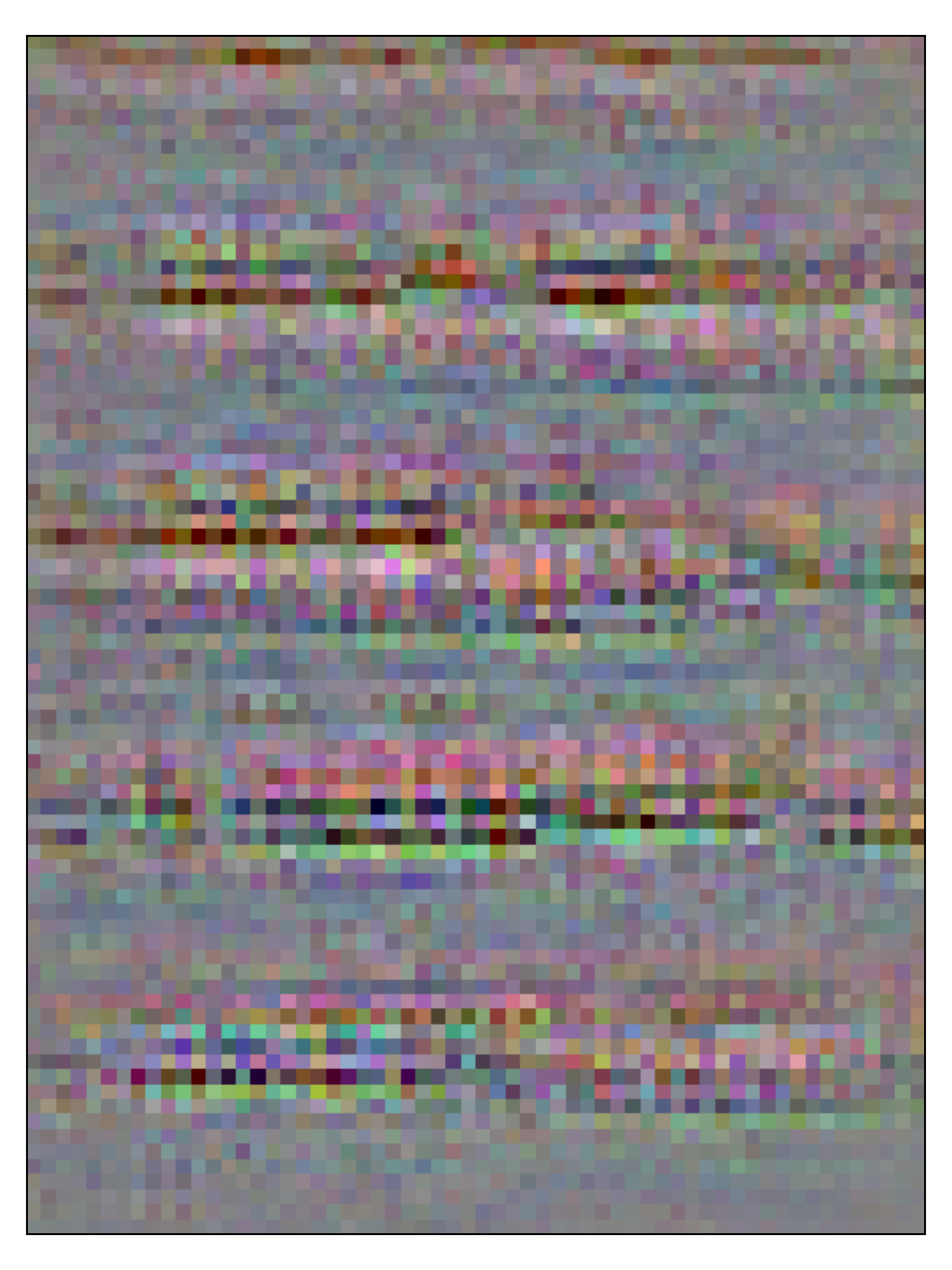}
			\includegraphics{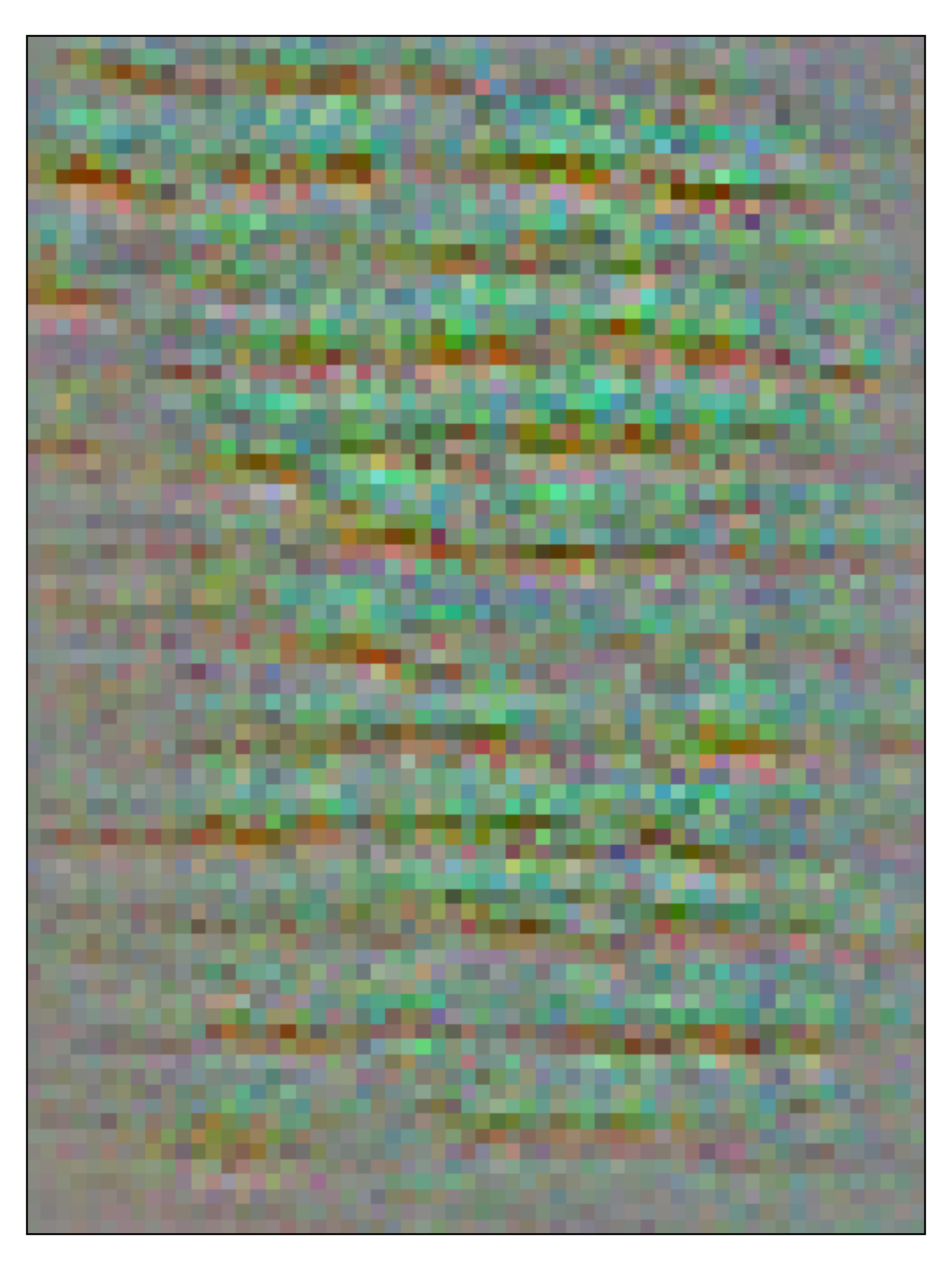}
			\includegraphics{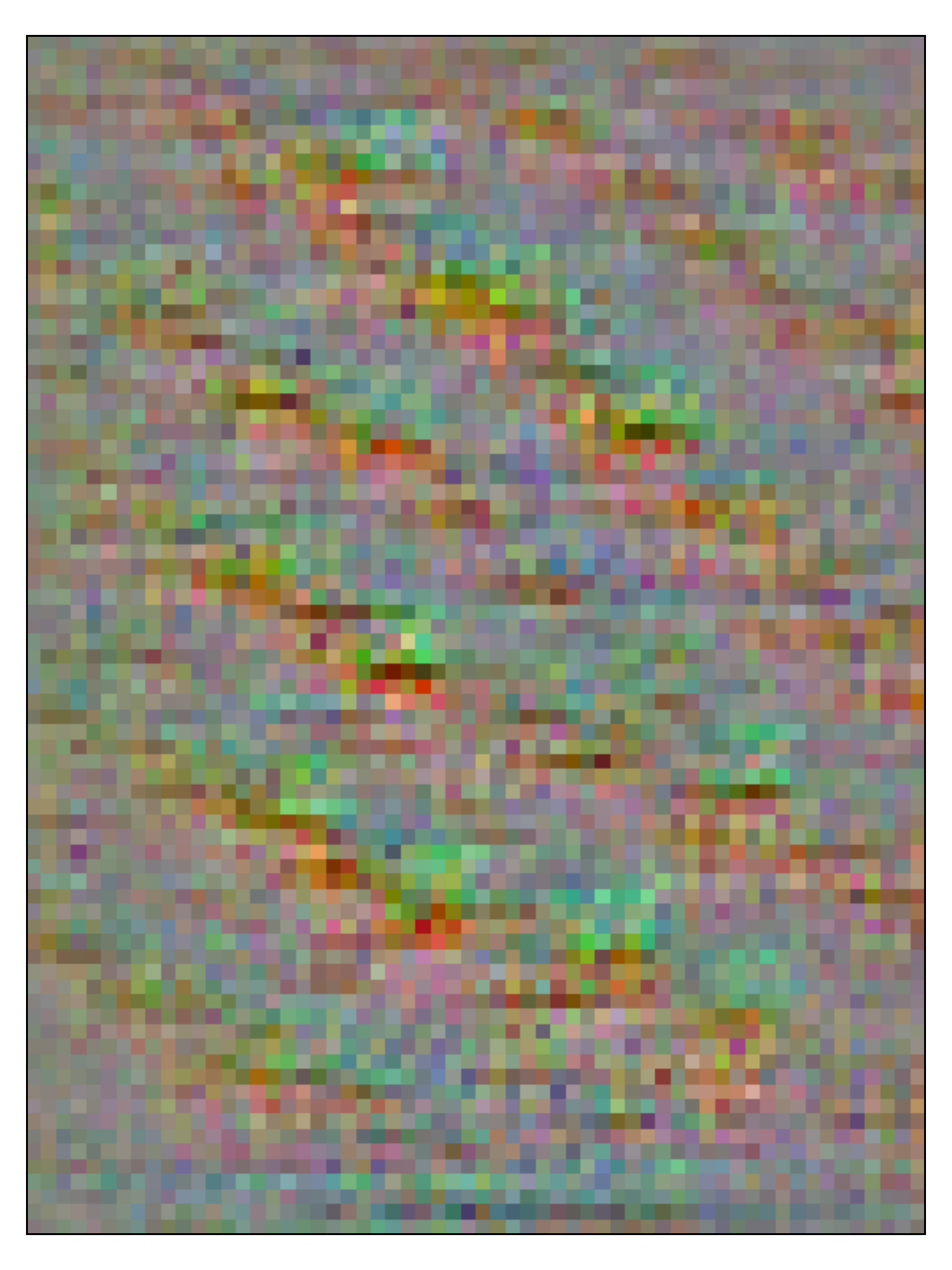}
			\includegraphics{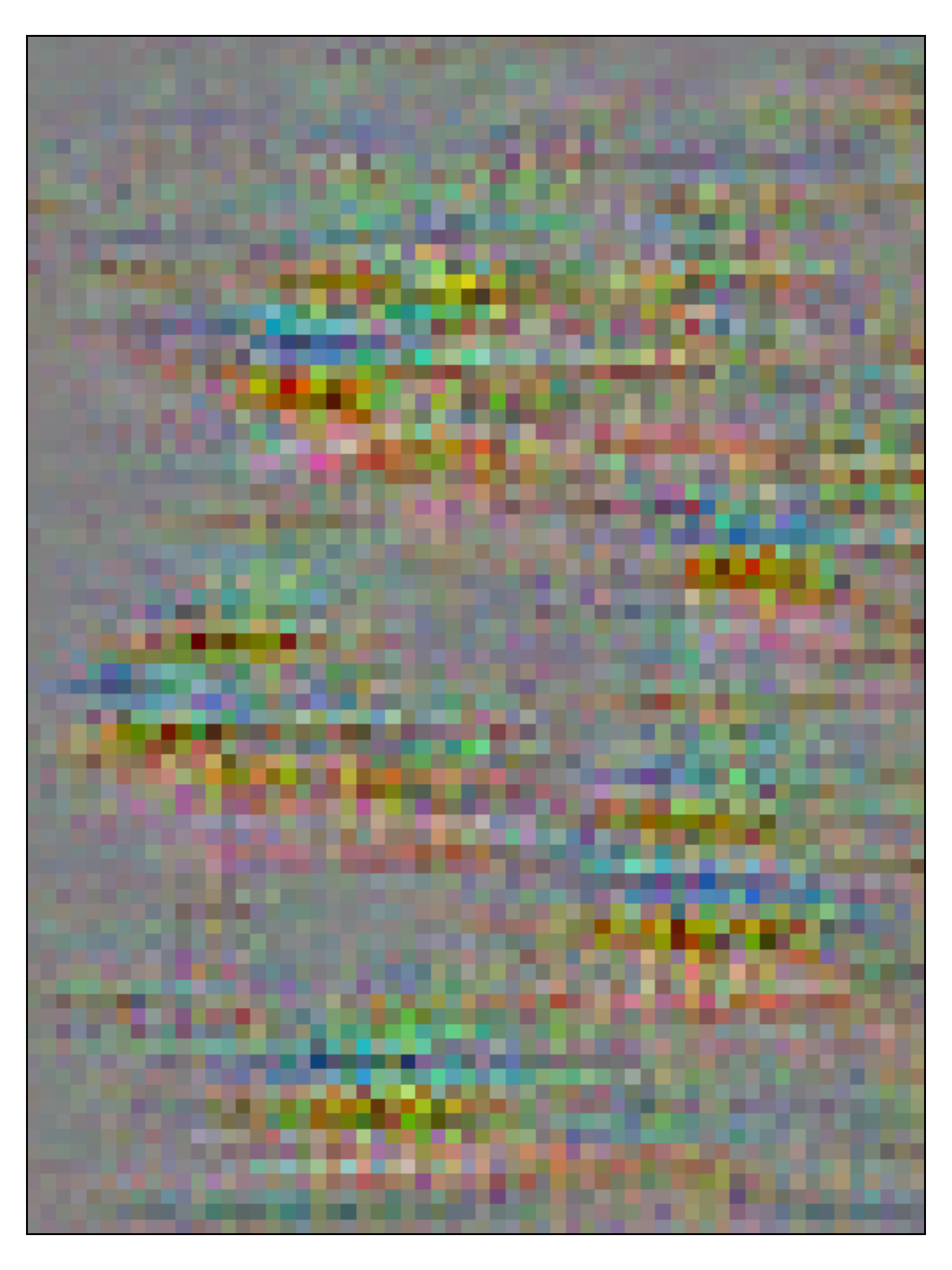}
			\includegraphics{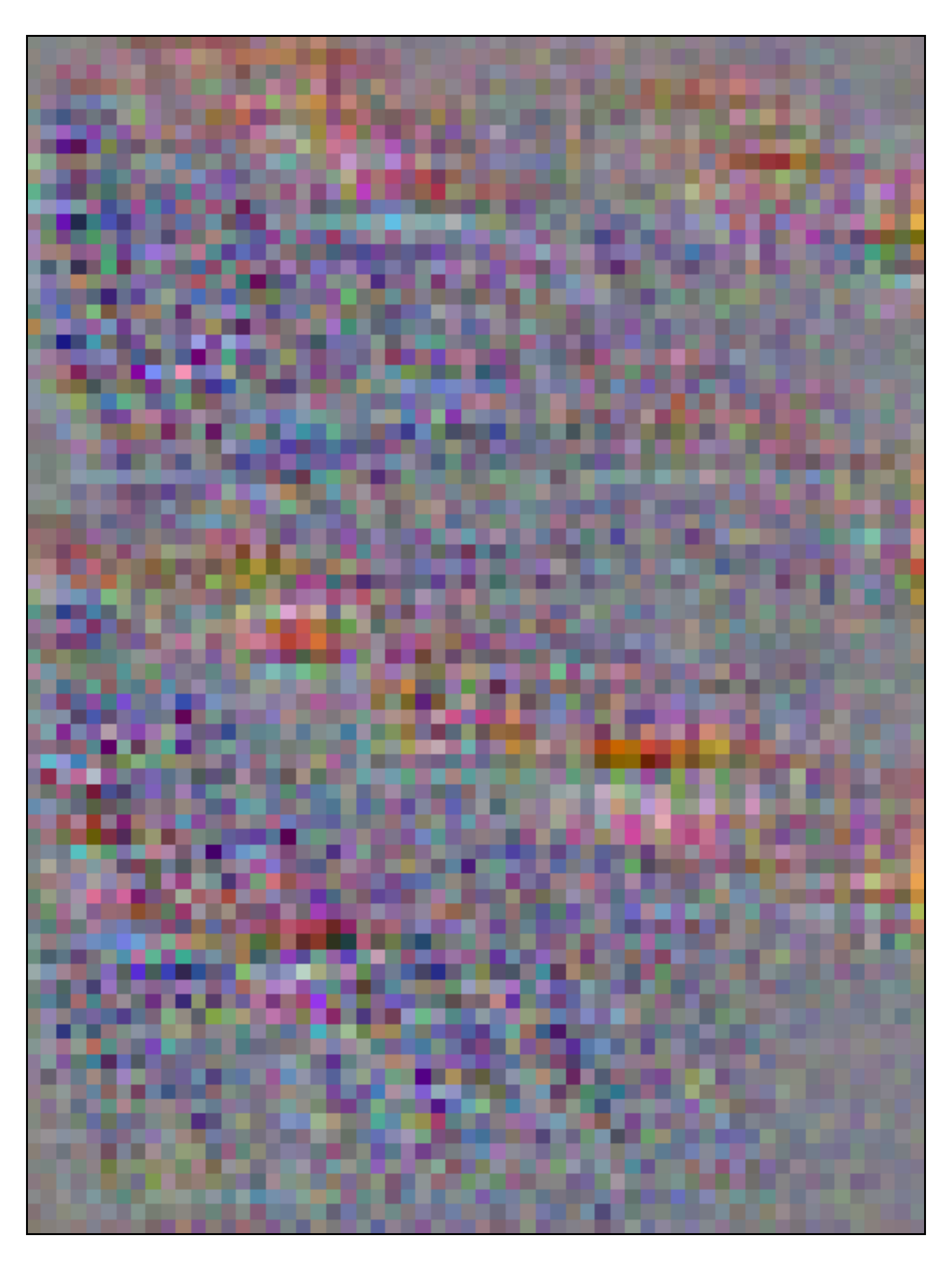}
			\includegraphics{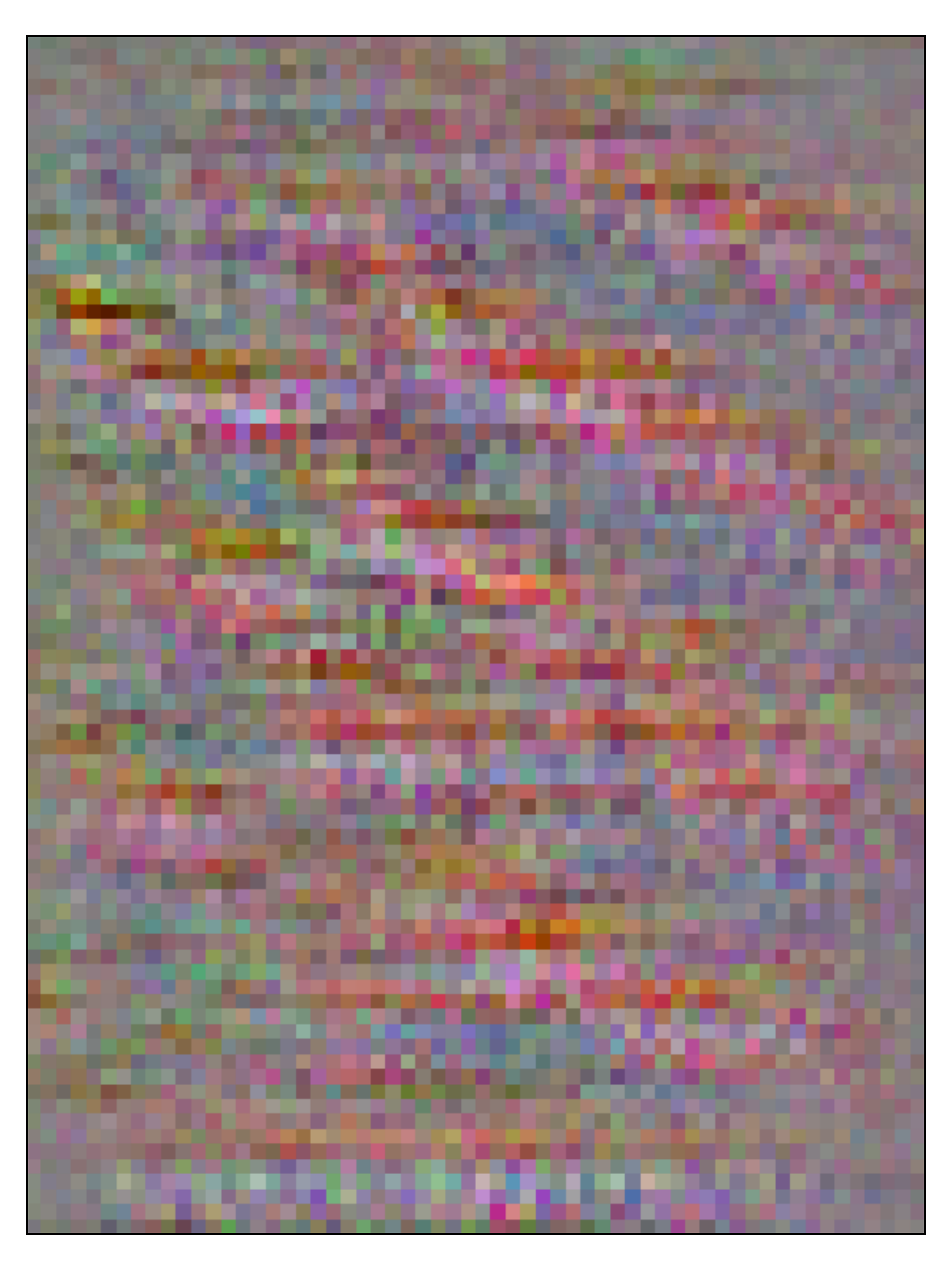}}
		\label{fig:kernel-repc}}
	\caption{\small Primary features learned by each layer of the trained speed reconstruction model (only a subset of kernels in each layer are shown here).}
	\label{fig:kernel-rep}
\end{figure*}

Interestingly, we see that the patterns or features learned by the model are more relevant to the macroscopic traffic speed states. For instance, the kernels in the first layer (Fig.~\ref{fig:kernel-repa}) detect discrete traffic states such as congestion (pale red color), free flow (purple color), slow moving traffic (red stripes) and vehicles in the transition regime (yellow). In the second layer (Fig.~\ref{fig:kernel-repb}), the kernels detect information propagation for  congested and free flow traffic as seen from the sloped green, red, and purple regions. The third layer (Fig.~\ref{fig:kernel-repc}) combines these primitive features and produces more realistic spatio-temporal patterns such as free flow regime, congested regime, transient dynamics (from free flow to congestion and vice-versa). This clearly depicts the model's ability to recognize different traffic phenomena, and the mechanisms by which this information is used to reconstruct the complete spatio-temporal speed maps.

\begin{figure*}[h!]
	\centering
	\subfloat[][Full trajectory]{\resizebox{0.235\textwidth}{!}{
			\includegraphics[width=0.235\textwidth,origin=c]{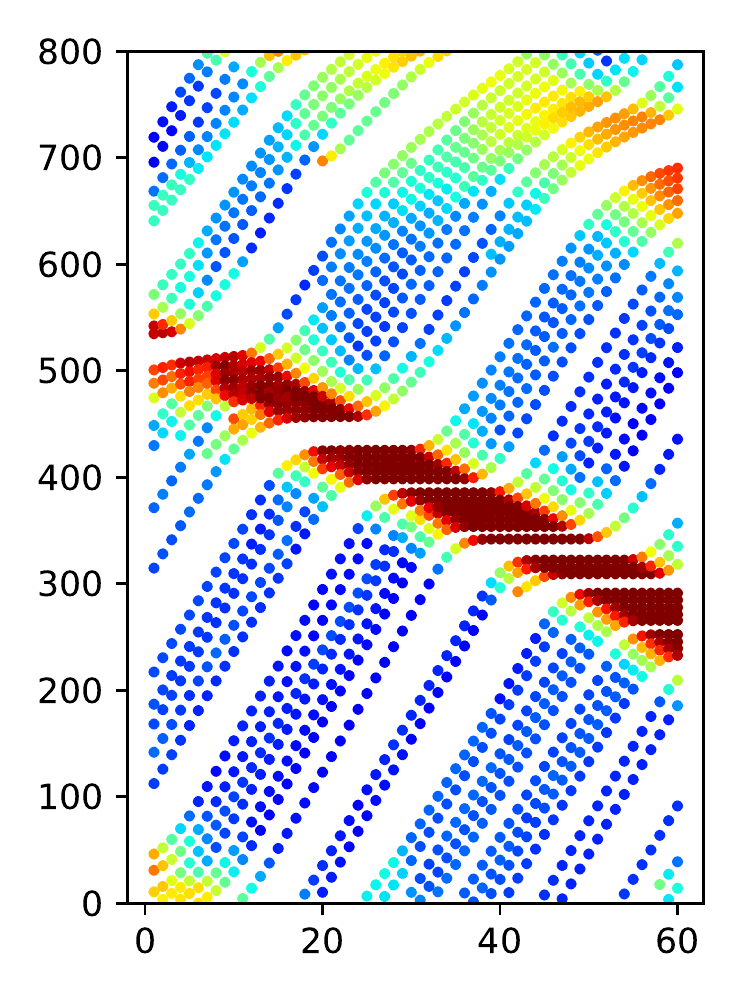}} 
		\label{fig:latent-repa}} 		
	\subfloat[][1\%]{\resizebox{0.1755\textwidth}{!}{
			\includegraphics[width=0.1755\textwidth,origin=c]{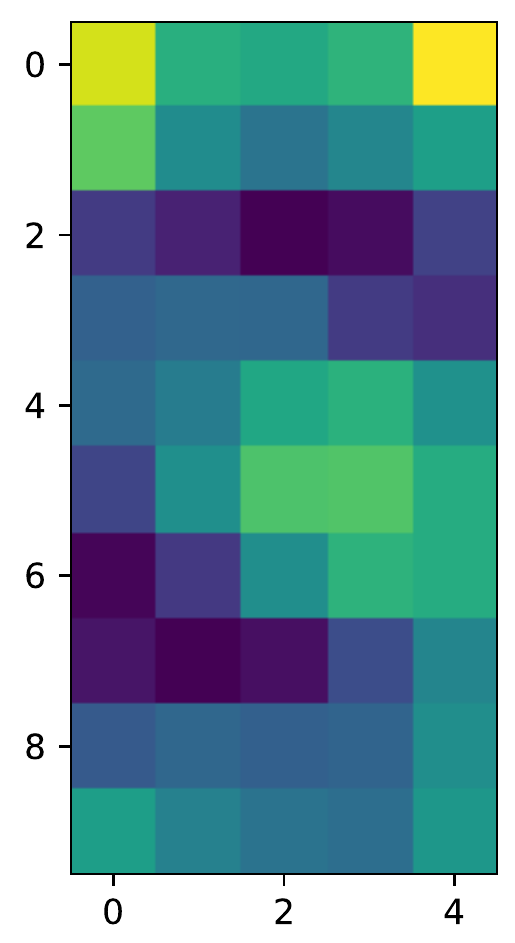}} 
		\label{fig:latent-repb}}
	\subfloat[][5\%]{\resizebox{0.175\textwidth}{!}{
			\includegraphics[width=0.175\textwidth,origin=c]{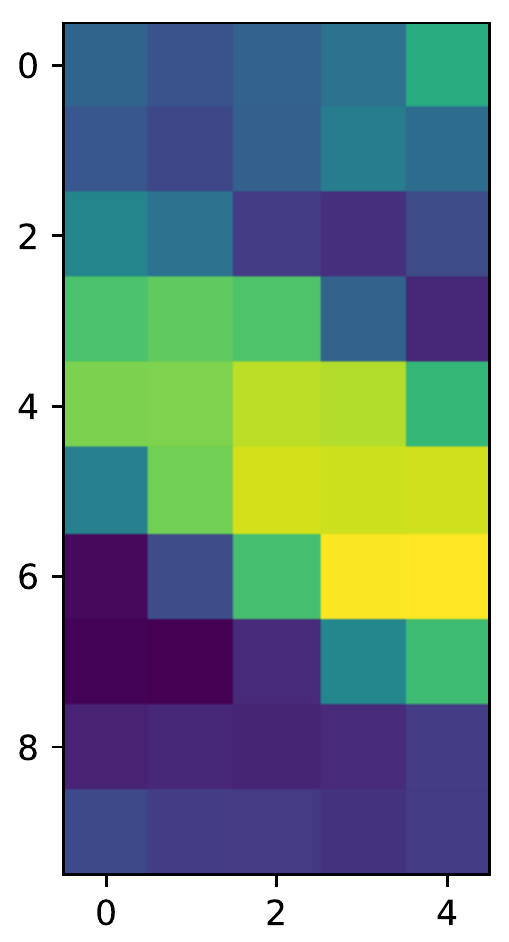}} 
		\label{fig:latent-repc}}
	\subfloat[][10\%]{\resizebox{0.1755\textwidth}{!}{
			\includegraphics[width=0.1755\textwidth,origin=c]{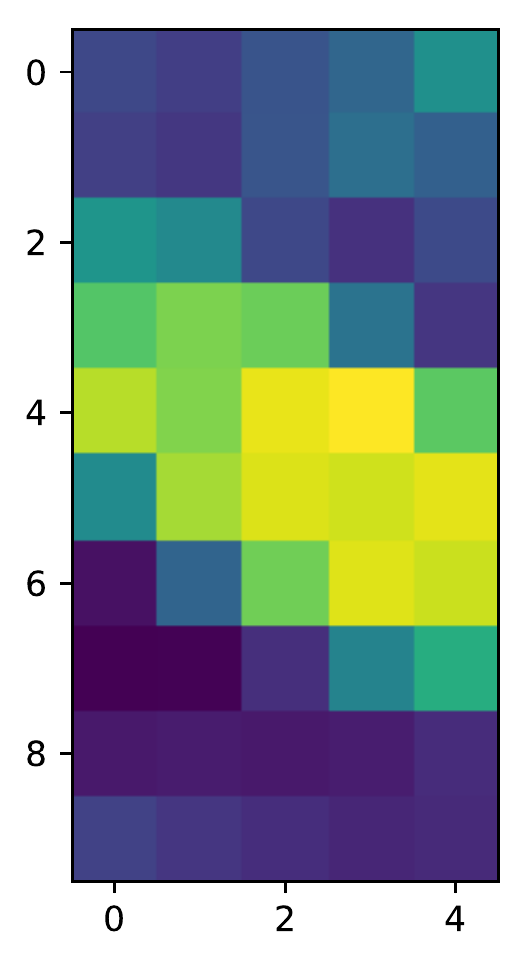}} 
		\label{fig:latent-repd}}
	\subfloat[][100\%]{\resizebox{0.22\textwidth}{!}{
			\includegraphics[width=0.22\textwidth,origin=c]{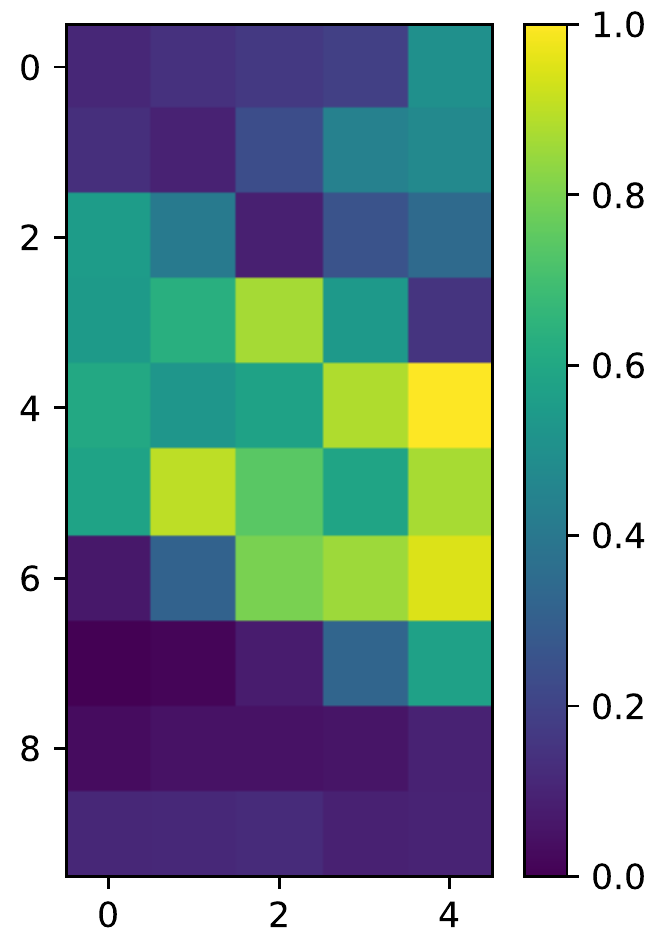}} 
		\label{fig:latent-repe}}
	\caption{\small Normalized latent space activations for different probe levels. To reduce the variance, the input trajectories are sampled 100 times and average activations are shown.}
	\label{fig:latent-rep}
\end{figure*}

To further understand the reconstruction abilities of the model, we investigate the latent space representation ($\mathbf{h}$) for different probe levels. Here, $\mathbf{h} \in \mathbb{R}^{(10 \times 5 \times 64)}$ contains 64 feature maps, each of which has a dimension of $(10 \times 5)$ (see Table~\ref{tab:model-arch}). We take the (normalized) sum of all these feature map activations in order to visualize them over a two-dimensional plane; this is shown in  Fig.~\ref{fig:latent-rep} for probe levels of 1\%, 5\%, 10\%, and 100\%.

Recall that $\mathbf{h}$ encodes the primary information from the sparse input vehicle trajectories, and the decoder model uses this encoded information to reconstruct the macroscopic speed states; i.e., the output from the decoder model only depends on $\mathbf{h}$. Assuming that the visualized activations in Fig.~\ref{fig:latent-rep} are a surrogate for the information contained in $\mathbf{h}$, we see similar patterns at 100\% and 5\% probe levels (in Fig.~\ref{fig:latent-repc} and Fig.~\ref{fig:latent-repe}), implying that the reconstructed speed states will also be the same. This explains why the estimation model performs well even at 5\% probe penetration rates.

\section{Conclusions}
\label{sec:conclusion}
In this paper, we address the problem of estimating dynamic traffic states (namely speeds) from limited probe vehicle data. We proposed a convolutional encoder-decoder neural network model to learn traffic speed dynamics from space-time diagrams. We illustrated this for a long road section and the results showed a sound and accurate reconstruction of macroscopic speed maps, even at 5\% probe penetration rate. The model captured various spatio-temporal traffic behaviors such as backward propagating shockwaves, free-flow regimes and transient dynamics, which existing estimation methods fail to reproduce at low probe penetration rates. This was further confirmed with an analysis of the model's reconstruction mechanism where we visually observed that the kernels in the initial layers of the CNN learned to identify discrete traffic states: free flow, congested, and transient states. Successive layers recognized more compound patterns such as stop-and-go traffic, free flow propagation, and transient dynamics. 

We also found that the model learned to reproduce dynamically varying shockwave patterns depending on the local traffic conditions, thus circumventing the constant wave speed assumption widely used in many of the existing estimation tools.  Furthermore, when validated against the NGSIM dataset, the neural network successfully identified all shockwaves, \emph{despite being trained using simulated data}. These insights attest to the viability of data-driven methods for real world applications. In the future, our efforts will continue along the lines of rigorous analysis of the model's reconstruction mechanism, as well as further applications to arterials with signalized intersections.


\section*{Acknowledgment}
This work was supported by the NYUAD Center for Interacting Urban Networks (CITIES), funded by Tamkeen under the NYUAD Research Institute Award CG001 and by the Swiss Re Institute under the Quantum Cities\textsuperscript{TM} initiative.

\appendix
\gdef\thesection{Appendix \Alph{section}}

	
	
	
\bibliographystyle{plainnat}
\bibliography{refs_R1.bib}
	
	
	
	
	
	

\end{document}